\documentclass[preprint,12pt]{elsarticle}

\usepackage{amsmath,amsthm, amssymb}

\newcommand{\dst}{\displaystyle}

\newcommand{\be}{\begin{equation}}
\newcommand{\ee}{\end{equation}}
\newcommand{\ba}{\begin{array}}
\newcommand{\ea}{\end{array}}

\newcommand{\bea}{\begin{eqnarray}}
\newcommand{\eea}{\end{eqnarray}}
\newcommand{\bma}{\begin{matrix}}
\newcommand{\ema}{\end{matrix}}
\newcommand{\bpm}{\begin{pmatrix}}
\newcommand{\epm}{\end{pmatrix}}
\newcommand{\nn}{\nonumber}
\newcommand{\half}{\frac{1}{2}}
\newcommand{\qter}{\frac{1}{4}}
\newcommand{\mc}{\mathcal}

\newcommand{\p}{\partial}

\newcommand{\rr}{\prime}
\newcommand{\ov}{\overline}
\newcommand{\wh}{\widehat}
\newcommand{\wt}{\widetilde}

\newcommand{\Psibar}{{\ov \Psi}{}}

\newcommand{\psibar}{{\ov \psi}{}}
\newcommand{\etabar}{{\ov \eta}}

\newcommand{\labar}{{\ov \la}{}}
\newcommand{\xibar}{{\ov \xi}}

\newcommand{\phibar}{\ov\phi{}}

\newcommand{\eps}{\varepsilon}
\newcommand{\ep}{\epsilon}

\newcommand{\al}{\alpha}

\newcommand{\la}{\lambda}
\newcommand{\da}{\delta}
\newcommand{\om}{\omega}
\newcommand{\Ga}{\Gamma}
\newcommand{\ga}{\gamma}
\newcommand{\La}{\Lambda}
\newcommand{\Si}{\Sigma}
\newcommand{\si}{\sigma}
\newcommand{\ta}{\theta}

\newcommand{\qrq}{\quad\Rightarrow\quad}

\newcommand{\Dbar}{\ov D}

\newcommand{\epbar}{{\ov\ep}}

\newcommand{\lra}{\leftrightarrow}

\newcommand{\pbar}{{\ov\p}}

\newcommand{\Abar}{{\ov A}{}}

\newcommand{\Gat}{\Ga\!{}_{_\triangle}}
\newcommand{\Qbar}{{\ov Q}}

\journal{Nuclear Physics B}

\begin{document}

\begin{frontmatter}

\title{\bf Dynamical supersymmetry\\ in maximally supersymmetric gauge theories}


\author{Dmitry V.~Belyaev}
\ead{belyaev@phys.ufl.edu}

\address{Institute for Fundamental Theory,
Department of Physics,\\ University of Florida,
Gainesville, FL 32611, USA}

\begin{abstract}
Maximally supersymmetric theories can be described by a single scalar superfield in light-cone superspace. When they are also (super)conformally invariant, they are uniquely specified by the form of the dynamical supersymmetry. We present an explicit derivation of the light-cone superspace form of the dynamical supersymmetry in the cases of ten- and four-dimensional super-Yang-Mills, and the three-dimensional Bagger-Lambert-Gustavsson theory, starting from the covariant formulation of these theories.
\end{abstract}


\end{frontmatter}


\numberwithin{equation}{section}

\section{Introduction}

Most known maximally supersymmetric theories are descendants of $d=10$ $N=1$ super-Yang-Mills (SYM) theory \cite{BSS,GSO} and $d=11$ $N=1$ supergravity \cite{CJS}. The $d=4$ $N=4$ SYM \cite{BSS} (with $16$ supercharges) and $d=4$ $N=8$ supergravity \cite{CJ} (with $32$ supercharges) are two best known examples. However, there are maximally supersymmetric theories that do not arise from the higher-dimensional parents, and the recently discovered $d=3$ $N=8$ Bagger-Lambert-Gustavsson (BLG) theory \cite{BL1,Gus1,BL2} is one of the ``exceptional'' cases. Nonetheless, all of these theories have something to do with $\mc{M}$-theory and all of them can be described in light-cone (LC) superspace \cite{BLN,ABR1,ABR2,ABKR,BKR,BN,BBKR}. The LC superspace form of these theories is quite interesting in its own right (pointing to some hidden simplicity of maximally supersymmetric theories), and it is also a convenient starting point for discussing quantum aspects of these theories \cite{SM,BLN2}.

The $d=4$ $N=4$ SYM and $d=3$ $N=8$ BLG theory are not only maximally supersymmetric (with $16$ supercharges), but also superconformally invariant.\footnote{
There exists a mass deformation of the BLG theory that preserves all 16 supersymmetries, but breaks conformal (super)symmetries \cite{mass1,mass2}. 
} 
The superconformal groups are, respectively, $PSU(2,2|4)$ and $OSp(2,2|8)$ \cite{Nahm}. They have $SU(4)$ and $SO(8)$ as their $R$-symmetry groups, respectively. In addition to 16 regular supersymmetries there are also 16 conformal supersymmetries. Although the two theories have very different covariant descriptions, they are quite similar in the LC superspace description. In particular, they are described by the same scalar superfield $\phi_a$ ($a$ is the gauge index) with $8+8$ degrees of freedom \cite{BLN,BN}. Interactions are governed by structure constants $f^{b c}{}_a$ and $f^{b c d}{}_a$, respectively.

In LC superspace, 16 regular supersymmetries split into $4+4$ kinematical supersymmetries ($q^m$ and $q_m$) and $4+4$ dynamical supersymmetries ($Q^m$ and $Q_m$). (Here $m$ is the fundamental $SU(4)$ index.) The distinction is simple: dynamical generators receive corrections from the interaction (they are $f$-dependent), while kinematical ones do not (they are $f$-independent) \cite{BBR}. A theory is fully specified when all dynamical generators (the Hamiltonian shift $\mc{P}^{-}$, the Lorentz boost $\mc{J}^{-}$, the dynamical conformal generator $\mc{K}^{-}$, dynamical supersymmetry generators $Q^m$ and dynamical conformal supersymmetry generators $S^m$) are given \cite{Dirac}. However, the superconformal groups are \emph{simple} \cite{Nahm} and this implies that once the dynamical supersymmetry generator $Q^m$ is known, other dynamical generators can be found by repeated commutation. The purpose of this paper is to show how one can derive the form of the dynamical supersymmetry in LC superspace from the covariant formulation of the theory.

The paper is organized as follows.
In Section~\ref{sec-LCS}, we give a short review of LC superspace. In Section~\ref{sec-alg}, we formulate our algorithm for deriving the LC superspace form of the dynamical supersymmetry transformation starting with the covariant formulation of a maximally supersymmetric gauge theory. In Section~\ref{sec-SYM}, we review the covariant formulation of the $d=10$ $N=1$ SYM which yields the $d=4$ $N=4$ SYM upon dimensional reduction. In Section~\ref{sec-dsSYM}, through the application of our algorithm, we derive the form of the dynamical supersymmetry in the SYM theories. Our result for the $d=4$ SYM matches the one in \cite{ABKR}, whereas our result for the $d=10$ SYM completes the one in \cite{ABR1}. Then we go through the same steps for the $d=3$ $N=8$ BLG theory. In Section~\ref{sec-BLG}, we review its covariant formulation, and in Section~\ref{sec-dsBLG}, we derive the form of the dynamical supersymmetry in the BLG theory. The result of our ``top-down'' derivation matches nicely with the result of the ``bottom-up'' approach advocated in \cite{BBKR}. Some technical details and clarifications are given in the appendices.

\section{LC superspace}
\label{sec-LCS}

Light-cone (LC) superspace is the usual superspace that makes kinematical supersymmetries manifest. For maximally supersymmetric theories (with 16 supercharges), the number of kinematical supersymmetries is 8 and so we need 8 Grassmann coordinates. These are conveniently chosen as four complex coordinates $\ta^m$ and their complex conjugates $(\ta^m)^\ast\equiv\ta_m$. (For more on complex conjugation see \ref{sec-cc}.) Kinematical supersymmetry generators are defined in the standard way \cite{BLN}~\footnote{
Every object with lower $SU(4)$ indices ($\ta_m$, $q_m$, $d_m$, $\chi_m$, $C_{m n}$, etc.) is accompanied with a bar in \cite{BLN}. We keep this bar implicit, and show it explicitly only when the indices are not shown. For example, $\zeta\bar{q}=\zeta^m q_m$. Note that the relation between the (barred) objects with lower indices and the result of complex conjugation of the corresponding (non-barred) objects with upper indices depends on the way one defines complex conjugation. With the definition used in this paper (see \ref{sec-cc}) one finds, for example, $(q^m)^\ast=-q_m$.
}
\bea
\label{defqm}
q^m=-\p^m+\frac{i}{\sqrt2}\ta^m\p^{+}, \quad
q_m=+\p_m-\frac{i}{\sqrt2}\ta_m\p^{+}\ ,
\eea
where
\bea
&& \p_m=\frac{\p}{\p\ta^m}, \quad
\p^m=\frac{\p}{\p\ta_m}, \quad
\{\p_m,\ta^n\}=\da_m^n, \quad
\{\p^m,\ta_n\}=\da^m_n \nn\\
&& \p^{+}=-\p_{-}, \quad 
\p_{-}=\frac{\p}{\p x^{-}}, \quad 
x^{-}=\frac{1}{\sqrt2}(x^0-x^3), \quad{}
[\p^{+},x^{-}]=-1\ . \qquad
\eea
Two kinematical supersymmetries anticommute into the translation along $x^{-}$
\bea
\label{qmqn}
\{q^m,q_n\}=+i\sqrt2\da^m_n\p^{+}\ .
\eea
Superspace covariant derivatives (``chiral derivatives'') are defined as follows
\bea
\label{defdm}
d^m=-\p^m-\frac{i}{\sqrt2}\ta^m\p^{+}, \quad
d_m=+\p_m+\frac{i}{\sqrt2}\ta_m\p^{+}\ .
\eea
They anticommute with $q^m$ and $q_m$, and the only nontrivial anticommutator is
\bea
\label{dmdn}
\{d^m,d_n\}=-i\sqrt2\da^m_n\p^{+}\ .
\eea
The central object in this LC superspace, for our purposes, is a scalar superfield $\phi$ satisfying the usual chirality condition
\bea
d^m\phi=0\ ,
\eea
as well as the following reality condition (the ``inside-out'' constraint)
\bea
\label{inout}
\phibar\equiv\phi^\ast=\frac{d_{[4]}}{2\p^{+2}}\phi, \quad
d_{[4]}\equiv\frac{1}{4!}\eps^{i j k l}d_{i j k l} \ ,
\eea
which can equivalently be stated as~\footnote{
We use the notation $d_{m n}=d_m d_n$, $d^{m n k}=d^m d^n d^k$, etc. These objects are antisymmetric in the $SU(4)$ indices because $\{d_m,d_n\}=0$ and $\{d^m,d^n\}=0$. Same applies to $\ta^{m n k}=\ta^m\ta^n\ta^k$, etc., thanks to $\{\ta^m,\ta^n\}=0$.
}
\bea
d_{m n}\phi=\half\eps_{m n p q}d^{p q}\phibar\ .
\eea
The superfield $\phi$ that satisfies the above two constraints has $8+8$ degrees of freedom which are given names through the following standard component expansion \cite{BLN}
\bea
\label{LCSF}
\phi(y)=\frac{1}{\p^{+}}A(y)+\frac{i}{\sqrt2}\ta^{m n}C_{m n}(y)
+\frac{1}{12}\ta^{m n p q}\eps_{m n p q}\p^{+}\Abar(y) \nn\\
+\frac{i}{\p^{+}}\ta^m\chi_m(y)+\frac{\sqrt2}{6}\ta^{m n p}\eps_{m n p q}\chi^q(y)\ . \hspace{50pt}
\eea
Here ``$y$'' refers to the shifted $x^{-}$ coordinate, $y^{-}=x^{-}-\frac{i}{\sqrt2}\ta^m\ta_m$, which in the usual way defines the dependence of a chiral superfield on the conjugated superspace coordinates ($\ta_m$).
The reality (``inside-out'') constraint requires that
\bea
(A)^\ast=\Abar, \quad (\chi_m)^\ast=\chi^m, \quad (C_{m n})^\ast=\half\eps^{m n p q}C_{p q}\equiv C^{m n}\ ,
\eea
so that there are $1+6+1$ independent real bosonic components and $4+4$ fermionic ones. The components can equivalently be defined via projection with covariant derivatives. We find that
\bea
\label{projphi}
\phi_|=\frac{1}{\p^{+}}A, \quad
d_m\phi_|=\frac{i}{\p^{+}}\chi_m, \quad
d_{m n}\phi_|=-i\sqrt2 C_{m n} \nn\\[5pt]
d_{m n k}\phi_|=-\sqrt2\eps_{m n k l}\chi^l, \quad
d_{m n k l}\phi_|=2\eps_{m n k l}\p^{+}\Abar \ , \quad
\eea
where the vertical bar indicates setting $\ta^m=\ta_m=0$. Similarly,
\bea
\label{projphibar}
\phibar_|=\frac{1}{\p^{+}}\Abar, \quad
d^m\phibar_|=-\frac{i}{\p^{+}}\chi^m, \quad
d^{m n}\phibar_|=-i\sqrt2 C^{m n} \nn\\[5pt]
d^{m n k}\phibar_|=\sqrt2\eps^{m n k l}\chi_l, \quad
d^{m n k l}\phibar_|=2\eps^{m n k l}\p^{+}A \ . \qquad
\eea
Note that $\phibar$ satisfies the antichirality condition $d_m\phibar=0$. Kinematical supersymmetries ($q$'s) lead to the following variation of $\phi$~\footnote{
We take $(\zeta^m)^\ast=\zeta_m$ which then requires the relative minus sign in (\ref{SFkinsusy}); see \ref{sec-cc} for more details.
}
\bea
\label{SFkinsusy}
\da_{\zeta\bar{q}}\phi=+\zeta^m q_m\phi, \quad
\da_{\bar{\zeta}q}\phi=-\zeta_m q^m\phi \ ,
\eea
where $\zeta$'s are (infinitesimal, anticommuting) supersymmetry parameters. 
The component supersymmetry transformations can be deduced from (\ref{SFkinsusy}) and we find
\bea
\label{ACkinsusy}
&& \da_{\zeta\bar{q}}A=i\zeta^m\chi_m, \quad
\da_{\bar\zeta q}A=0 \nn\\[5pt]
&& \da_{\zeta\bar{q}}C_{m n}=-i\eps_{m n k l}\zeta^k\chi^l, \quad
\da_{\bar\zeta q}C_{m n}=-i(\zeta_m\chi_n-\zeta_n\chi_m) \ .
\eea
We will use these transformations to deduce the embedding of component fields in the covariant formulation of a maximally supersymmetric gauge theory into the LC superfield $\phi_a$. We will also find how dynamical supersymmetry transformations ($Q$'s) become realized on the superfield $\phi_a$. This realization will be nonlinear in the superfields \cite{ABKR,BBR}, with the nonlinear part depending on the structure constants $f$.~\footnote{
The $f$-independent part of a dynamical generator is linear in $\phi_a$ and as such is model-independent (but it does depend on the dimension of spacetime).
}

\section{The algorithm}
\label{sec-alg}

Starting from the (Lorentz-)covariant formulation of the $d=10$ $N=1$ SYM and the $d=3$ $N=8$ BLG theory, we will go through the following steps to arrive at the form of the dynamical supersymmetry in LC superspace:
\begin{itemize}
\item[1)]
impose the LC gauge;
\item[2)]
use equations of motion to solve for dependent field components;
\item[3)]
find supersymmetry transformations of independent bosonic components (taking into account compensating gauge transformations needed to stay in the LC gauge);
\item[4)]
match the kinematical part of the supersymmetry transformations onto (\ref{ACkinsusy}) to identify $A$, $C_{m n}$ and $\chi_m$;
\item[5)]
use the dynamical part of the supersymmetry transformation of $A$ to guess the corresponding transformation of $\phi$;
\item[6)]
make consistency checks to verify the guess.
\end{itemize}
We will see that splitting supersymmetry transformations in the LC gauge into the kinematical and dynamical parts is done quite easily. The lifting of the component transformation of $A$ to the superfield transformation of $\phi$ (recall that $A=\p^{+}\phi_|$) is not unique, however, and one has to use additional arguments to narrow down the possibilities and make consistency checks to get at the final answer. 

This algorithm requires only the knowledge of supersymmetry transformations and equations of motion in the covariant formulation. In particular, we will never need to know the Lagrangian of the theory. This allows us to stay with the structure constans $f^{b c}{}_a$ and $f^{b c d}{}_a$ as they follow from the algebra. (For the Lagrangian formulation one has to introduce, in addition, a metric for the gauge indices.) The BLG theory was introduced in this fashion in \cite{BL2}, and we will mimic their approach for the SYM case as well.

\newpage
\section{The $d=10$ $N=1$ SYM}
\label{sec-SYM}

To derive the $d=10$ $N=1$ super-Yang-Mills (SYM) theory \cite{BSS,GSO} from scratch, we will follow \cite{BL2} and start with a Lie group that has generators $T^a$ satisfying the following commutation relations~\footnote{
Indices inside the square brackets are antisymmetrized ``with strength one.'' E.g. $A^{[a b]}=\half(A^{a b}-A^{b a})$.
}
\bea
[T^b,T^c]=f^{b c}{}_a T^a, \quad
f^{b c}{}_a=f^{[b c]}{}_a \ .
\eea
A covariant object $X_a$ (in the adjoint representation of the gauge group) transforms under an infinitesimal gauge transformation with parameters $\om_a$ as follows
\bea
\da_\om X_a=f^{b c}{}_a X_b\om_c \ .
\eea
In order to turn its spacetime derivative $\p_M X_a$, ($M=0,\dots,9$), into a covariant object, with $\om_a$ being local ($\p_M\om_a\neq 0$), we need to introduce a gauge field $A_{M a}$ that transforms as follows
\bea
\label{AgtrSYM}
\da_\om A_{M a}=\p_M\om_a+f^{b c}{}_a A_{M b}\om_c \equiv D_M\om_a \ .
\eea
The covariant derivative of $X_a$ is then defined as
\bea
D_M X_a=\p_M X_a+f^{b c}{}_a A_{M b}X_c \ .
\eea
The commutator of two covariant derivatives defines the field-strength $F_{M N a}$
\bea
[D_M,D_N]X_a &=& f^{b c}{}_a F_{M N b}X_c \nn\\[3pt]
F_{M N a} &=& \p_M A_{N a}-\p_N A_{M a}+f^{b c}{}_a A_{M b}A_{N c} \ ,
\eea
which also transforms covariantly. We note that in deriving this expression for $F_{M N a}$ (as well as in the proof of covariance for both $D_M X_a$ and $F_{M N a}$) one has to use the Jacobi identity for the structure constants
\bea
\label{Jacobi}
f^{[b c}{}_g f^{d]g}{}_a=0 \ .
\eea

In $d=10$, the gauge field $A_M$ has 8 on-shell degrees of freedom, just as a 32-component Majorana-Weyl spinor $\la$. The pair $(A_{M a},\la_a)$ forms an (on-shell) supersymmetry multiplet with the following supersymmetry transformations
\bea
\da_\ep A_{M a} &=& i\epbar\Ga_M\la_a \nn\\[3pt]
\da_\ep\la_a &=& \half\Ga^{M N}\ep F_{M N a} \ ,
\eea
where $\ep$ is the supersymmetry parameter (also a 32-component Majorana-Weyl spinor). Our conventions are such that ($I_{32}$ is the $32\times32$ unit matrix)
\bea
\label{GaGaEta}
\{\Ga_M,\Ga_N\}=2\eta_{M N}I_{32}, \quad{} [\Ga^M,\Ga^N]=2\Ga^{M N}, \quad \eta_{M N}=(-+\dots+) \ ,
\eea
the bar stands for Majorana conjugation ($C$ is the charge conjugation matrix)
\bea
\epbar\equiv\ep^T C, \quad C^T=-C, \quad C\Ga_M C^{-1}=-(\Ga_M)^T \ ,
\eea
while the Majorana and Weyl conditions are~\footnote{
The dagger $\dagger$ operation is the combination of transposition and complex conjugation, $\ep^\dagger=(\ep^\ast)^T$. We prefer not to call it ``Hermitian conjugation'' for reasons discussed in \ref{sec-cc}. Note also that we use the notation ``$\Ga_\ast$'' instead of ``$\Ga_{11}$''.
}
\bea
\label{MW}
&& \ep^T C=\ep^\dagger\Ga_0, \quad \la_a^T C=\la_a^\dagger\Ga_0 \nn\\
&& \Ga_\ast\ep=\ep, \quad \Ga_\ast\la_a=\la_a; \quad \Ga_\ast\equiv\Ga_0\Ga_1\dots\Ga_9 \ .
\eea
For the commutator of two supersymmetry transformations we then find (see \ref{sec-Fierz} for useful identities)
\bea
[\da_{\ep_1},\da_{\ep_2}]A_{M a} &=& v^N F_{N M a} \nn\\[5pt]
[\da_{\ep_1},\da_{\ep_2}]\la_a &=& v^N D_N\la_a \nn\\[5pt]
&&\hspace{-40pt}
+\Big\{
\frac{7}{8}i(\epbar_2\Ga^N\ep_1)\Ga_N
-\frac{i}{16\cdot5!}(\epbar_2\Ga^{K L P Q R}\ep_1)\Ga_{K L P Q R}\Big\}E_a(\la) \ , \qquad
\eea
where we defined
\bea
v^M \equiv -2i(\epbar_2\Ga^M\ep_1), \quad
E_a(\la) \equiv \Ga^M D_M\la_a \ .
\eea
Taking $E_a(\la)=0$ to be the equation of motion for $\la_a$, we find that, on-shell,
\bea
[\da_{\ep_1},\da_{\ep_2}]A_{M a} &=& v^N\p_N A_{M a}+D_M\om_a, \quad \om_a \equiv -v^N A_{N a} \nn\\[5pt]
[\da_{\ep_1},\da_{\ep_2}]\la_a &=& v^N\p_N A_{M a}+f^{b c}{}_a\la_b\om_c \ ,
\eea
so that the commutator of two supersymmetry transformations yields a translation and a gauge transformation (after the fermionic equation of motion has been used). Equations of motion should close under supersymmetry, and this allows us to derive the equation of motion for $A_{M a}$ from the one for $\la_a$ by supervariation. We find that
\bea
\da_\ep E_a(\la)=\Ga^N\ep E_{N a}(A), \quad
E_{N a}(A) \equiv D^M F_{M N a}+\frac{i}{2}f^{b c}{}_a\labar_b\Ga_N\la_c \ ,
\eea
so that $E_{N a}(A)=0$ is the bosonic equation of motion. This fully defines the $d=10$ $N=1$ SYM theory (and then $d=4$ $N=4$ SYM by dimensional reduction \cite{BSS,GSO}), and we now have all we need to apply the algorithm of Section \ref{sec-alg}.

\section{Dynamical supersymmetry in $d=10$ $N=1$ SYM}
\label{sec-dsSYM}

In this section, we will apply the algorithm of Section \ref{sec-alg} to $d=10$ $N=1$ SYM, arriving at the superfield form for dynamical supersymmetry transformations. The result for SYM theories in $d<10$ (and $d=4$ $N=4$ SYM in particular) will then straightforwardly follow by dimensional reduction.

\subsection{LC coordinates and LC gauge}
\label{sec-step1}

First, we introduce LC coordinates by defining \cite{BLN}
\bea
\label{LCx}
x^{\pm}=\frac{1}{\sqrt2}(x^0\pm x^3) \ .
\eea
Accordingly, we interpret the 10-dimensional vector index $M=0,1,\dots,9$ as having ``$+$'' and ``$-$'' as two of its entries: $M=(+,-,I)$. The 8-dimensional index $I$ is further split as follows
\bea
I=(s,I^\rr), \quad s=1,2, \quad I^\rr=\wh I+3, \quad \wh I=1,2,3,4,5,6 \ .
\eea
Starting with (\ref{GaGaEta}), we find that the metric tensor becomes off-diagonal with
\bea
\eta^{+-}=\eta^{-+}=-1, \quad \eta^{++}=\eta^{--}=0, \quad \eta^{I J}=\da^{I J} \ ,
\eea
and that the gamma matrix algebra becomes
\bea
\Ga^{+}\Ga^{-}+\Ga^{-}\Ga^{+}=-2 I_{32}, \quad
\Ga^{+}\Ga^{+}=\Ga^{-}\Ga^{-}=0, \quad
\{\Ga^I,\Ga^J\}=2\da^{I J}I_{32} \ .
\eea
We define the following projectors \cite{BLN}
\bea
\label{P+P-}
P_{+}=-\half\Ga_{+}\Ga_{-}=-\half\Ga^{-}\Ga^{+}, \quad
P_{-}=-\half\Ga_{-}\Ga_{+}=-\half\Ga^{+}\Ga^{-}\ ,
\eea
which satisfy all the required properties
\bea
P_{+}+P_{-}=I_{32}, \quad P_{\pm}P_{\pm}=P_{\pm}, \quad P_{\pm}P_{\mp}=0 \ .
\eea
We use them to (uniquely) decompose every 32-component spinor as follows
\bea
\ep=\ep_{+}+\ep_{-}, \quad \ep_{\pm}\equiv P_{\pm}\ep \ .
\eea
For the Majorana-conjugated spinor, we then find
\bea
\epbar=\epbar_{+}+\epbar_{-}, \quad \epbar_{\pm}\equiv (\ep_{\pm})^T C=\epbar P_{\mp} \ .
\eea
After these preparations, we choose our LC gauge as follows~\footnote{
Note that the LC gauge does not fix the gauge redundancy uniquely. The residual gauge transformations are those with parameter $\om_a$ satisfying $\p^{+}\om_a=0$.
}
\bea
\label{LCg}
\boxed{
\text{LC gauge: } \quad A_{-a}=0 \ .
}
\eea
Noting that $A_{-}=-A^{+}$ and $\p_{-}=-\p^{+}$, we then find that
\bea
A^{+}_a=0, \quad F_{M-a}=\p^{+}A_{M a} \ .
\eea

\subsection{Dependent field components}
\label{sec-step2}

Let us now use the equations of motion
\bea
&& E_a(\la) \equiv \Ga^M D_M\la_a =0 \nn\\
&& E_{N a}(A) \equiv D^M F_{M N a}+\frac{i}{2}f^{b c}{}_a\labar_b\Ga_N\la_c=0
\eea
to solve for field components that become dependent in the LC gauge. To do so, we separate the parts of the equations which do not involve the dynamical ``time derivative'' $\p^{-}$, and then use the freedom of inverting the nondynamical derivative $\p^{+}$. Hitting $E_a(\la)$ with $\Ga_{-}$ and using that $\Ga_{-}\Ga_{-}=0$ and $\Ga_{-}\Ga_{+}=-2P_{-}$ yields
\bea
-2D^{+}\la_{a-}+\Ga_{-}\Ga^I D_I\la_{a+}=0 \ .
\eea
In the LC gauge, $D^{+}=\p^{+}$ and we can easily ``solve'' this equation to find that~\footnote{
We have $\p^{+}=-\frac{\p}{\p x^{-}}$ and therefore $\frac{1}{\p^{+}}=-\int d x^{-}$.
}
\bea
\label{lasol}
\boxed{
\la_{a-}=\frac{1}{2\p^{+}}(\Ga_{-}\Ga^I D_I\la_{a+}), \quad
D_I\la_{a+}=\p_I\la_{a+}+f^{b c}{}_a A_{I b}\la_{c+} \ .
}
\eea
Turning to the bosonic equation of motion, we choose $N=$``$-$'' in $E_{N a}(A)$, which gives
\bea
\p^{+}F_{+-a}+D^I F_{I-a}+\frac{i}{2}f^{b c}{}_a\labar_b\Ga_{-}\la_c=0 \ .
\eea
Using that $F_{+-a}=\p^{+}A_{+a}$ then yields
\bea
\label{Aplus}
A_{+a}=-\frac{1}{\p^{+2}}(D^I F_{I-a}+\frac{i}{2}f^{b c}{}_a\labar_b\Ga_{-}\la_c) \ .
\eea
In the following, only the solution for $\la_{a-}$ will be used. It is also essential that $A_{I a}$ are all independent components.

\subsection{Modified supersymmetry transformations}
\label{sec-step3}

For supersymmetry transformations to be compatible with the LC gauge, we have to modify them by combining with appropriate compensating gauge transformations. We therefore define
\bea
\da_\ep^\rr A_{M a} &=& i\epbar\Ga_M\la+D_M\om_a \nn\\
\da_\ep^\rr \la_a &=& \half\Ga^{M N}\ep F_{M N a}+f^{b c}{}_a\la_b\om_c \ .
\eea
Compatibility with the LC gauge (\ref{LCg}) then requires
\bea
\da_\ep^\rr A_{-a} = i\epbar\Ga_{-}\la_a+\p_{-}\om_a=0 \ .
\eea
Using that $\p_{-}=-\p^{+}$, $P_{+}\Ga_{-}=0$ and $\Ga_{-}P_{-}=0$, we find
\bea
\label{omcomp}
\boxed{
\om_a=\frac{i}{\p^{+}}(\epbar_{+}\Ga_{-}\la_{a+}) \ .
}
\eea
The (modified) supersymmetry transformations for the independent bosonic fields then are
\bea
\da_\ep^\rr A_{I a}=i\epbar_{-}\Ga_I\la_{a+}+i\epbar_{+}\Ga_{I}\la_{a-}+D_I\om_a \ ,
\eea
where we should use (\ref{lasol}) and (\ref{omcomp}) to replace $\la_{a-}$ and $\om_a$. Noting that
\bea
D_I\om_a=\p_I\om_a+f^{b c}{}_a A_{I b}\om_c
=\frac{1}{\p^{+}}D_I(\p^{+}\om_a)+f^{b c}{}_a\frac{1}{\p^{+}}(\p^{+}A_{I b}\cdot\om_c) \ ,
\eea
and separating $\ep_{+}$ from $\ep_{-}$ transformations, we find
\bea
\label{kinsusyA}
\boxed{
\da^\rr_{\ep_{-}} A_{I a}=i\epbar_{-}\Ga_I\la_{a+}
}
\eea
and
\bea
\da^\rr_{\ep_{+}} A_{I a} &=& \frac{i}{2\p^{+}}(\epbar_{+}\Ga_I\Ga_{-}\Ga^J D_J\la_{a+})
+\frac{i}{\p^{+}}(\epbar_{+}\Ga_{-}D_I\la_{a+}) \nn\\[3pt]
&+& f^{b c}{}_a\frac{1}{\p^{+}}(\p^{+}A_{I b}\cdot\om_c) \ .
\eea
Using that $\Ga_I\Ga_{-}=-\Ga_{-}\Ga_I$ and $-\Ga_I\Ga^J=\Ga^J\Ga_I-2\da_I^J I_{32}$, this simplifies to
\bea
\label{dynsusyA}
\boxed{
\da^\rr_{\ep_{+}} A_{I a}=\frac{i}{2\p^{+}}(\epbar_{+}\Ga_{-}\Ga^J D_J\Ga_I\la_{a+})
+f^{b c}{}_a\frac{1}{\p^{+}}(\p^{+}A_{I b}\cdot\om_c) \ .
}
\eea
We observe that $\ep_{+}$ transformations involve both $1/\p^{+}$ and $f^{b c}{}_a$, whereas $\ep_{-}$ transformations involve neither $1/\p^{+}$ nor $f^{b c}{}_a$. We conclude, therefore, that $\ep_{-}$ transformations correspond to kinematical supersymmetry, whereas $\ep_{+}$ transformations are dynamical supersymmetry transformations.

\subsection{Identifying superfield components}
\label{sec-step4}

We found that, in the LC gauge (\ref{LCg}), the independent fields in the $d=10$ $N=1$ gauge multiplet $(A_{M a},\la_a)$ are given by $(A_{I a},\la_{a+})$. These independent fields must be in one-to-one correspondence with the components of the LC superfield $\phi_a$ as given in (\ref{LCSF}).
In order to identify superfield components, we need to find a map between $(A_{I a},\la_{a+})$ and $(A,C_{m n},\chi_m)_a$ that brings kinematical supersymmetry transformations (\ref{kinsusyA}) to the form of (\ref{ACkinsusy}). 
This task is facilitated by choosing a convenient representation for gamma matrices, and in this choice we will follow closely \cite{BLN} (modulo adjusting their representation to our conventions). We take (see \ref{sec-gm6} and \ref{sec-gm11} for the explicit form of $\ga$'s and $\Si$'s)
\bea
\Ga^0=i\ga^0\otimes I_8, \quad \Ga^3=i\ga^3\otimes I_8; \quad
\Ga^1=i\ga^1\otimes I_8, \quad \Ga^2=i\ga^2\otimes I_8 \ ,
\eea
and (note that $I=1,2,I^\rr$; $I^\rr=\wh I+3$; $\wh I=1,2,3,4,5,6$; $m=1,2,3,4$)
\bea
\Ga^{I^\rr}=i\ga_5\otimes\wh\Ga^{I^\rr-3}, \quad
\wh\Ga^{\wh I}=\bpm 0 & \Si^{\wh I m n} \\ \Si_{\wh I m n} & 0 \epm; \quad
I_8=\bpm \da^m_n & 0 \\ 0 & \da_m^n \epm \ .
\eea
In addition, we have
\bea
\Ga_\star \equiv \Ga_0\Ga_1\dots\Ga_9=\ga_5\otimes\bpm I_4 & 0 \\ 0 & -I_4 \epm, \quad
C=i C_4\otimes\bpm 0 & I_4 \\ I_4 & 0 \epm \ ,
\eea
and the projectors (\ref{P+P-}) become
\bea
P_{\pm}=p_{\pm}\otimes I_8; \quad p_{+}=\half\ga^{-}\ga^{+}, \quad p_{-}=\half\ga^{+}\ga^{-}; \quad
\ga^{\pm}=\frac{1}{\sqrt2}(\ga^0\pm\ga^3) \ . \quad
\eea
We will write the 32-component spinors in accordance with the ``$\otimes$'' structure of the gamma matrices,
\bea
\ep_{-}=\bpm \ep_{-}^m \\ \ep_{m-} \epm, \quad 
\ep_{+}=\bpm \ep_{+}^m \\ \ep_{m+} \epm, \quad 
\la_{a+}=\bpm \la_{a+}^m \\ \la_{m a+} \epm \ ,
\eea
where 
\bea
\label{4dproj}
&& \ep^m_{-}=p_{-}\ep^m_{-}, \quad \ep^m_{+}=p_{+}\ep^m_{+}, \quad \la^m_{a+}=p_{+}\la^m_{a+} \nn\\
&& \ep_{m-}=p_{-}\ep_{m-}, \quad \ep_{m+}=p_{+}\ep_{m+}, \quad \la_{m a+}=p_{+}\la_{m a+}
\eea
are 4-component spinors which (implicitly) carry indices on which $\ga$'s can act. The Majorana and Weyl conditions (\ref{MW}) reduce to
\bea
\label{4dMW}
&& (\ep^m_{-})^\ast=B_4\ep_{m-}, \quad
(\ep^m_{+})^\ast=B_4\ep_{m+}, \quad
(\la^m_{a+})^\ast=B_4\la_{m a+} \nn\\[5pt]
&& (\ep_{m-})^\ast=B_4\ep^m_{-}, \quad
(\ep_{m+})^\ast=B_4\ep^m_{+}, \quad
(\la_{m a+})^\ast=B_4\la^m_{a+} \nn\\[5pt]
&& \ga_5\ep^m_{-}=\ep^m_{-}; \quad
\ga_5\ep^m_{+}=\ep^m_{+}; \quad
\ga_5\la^m_{a+}=\la^m_{a+} \nn\\[5pt]
&& \ga_5\ep_{m-}=-\ep_{m-}, \quad
\ga_5\ep_{m+}=-\ep_{m+}, \quad
\ga_5\la_{m a+}=-\la_{m a+} \ , \qquad
\eea
where $B_4=C_4\ga_0$. It helps to write the $4\times4$ matrices entering (\ref{4dproj}) and (\ref{4dMW}) explicitly:
\bea
\ga_5=\bpm - & & & \\ & - & & \\ & & + & \\ & & & + \epm, \quad
B_4=\bpm & & & - \\ & & + & \\ & + & & \\ - & & & \epm
\nn\\[5pt]
p_{+}=\bpm 0 & & & \\ & + & & \\ & & + & \\ & & & 0 \epm, \quad
p_{-}=\bpm + & & & \\ & 0 & & \\ & & 0 & \\ & & & + \epm \ .
\eea
It is then easy to see that only one component in each 4-component spinor is nonzero. We will give names to these components by writing
\bea
\ep^m_{-}=\bpm 0 \\ 0 \\ 0 \\ \al^m \epm, \quad
\ep_{m-}=\bpm -\al_m \\ 0 \\ 0 \\ 0 \epm; \quad
\ep^m_{+}=\bpm 0 \\ 0 \\ \beta^m \\ 0 \epm, \quad
\ep_{m+}=\bpm 0 \\ \beta_m \\ 0 \\ 0 \epm
\eea
and
\bea
\label{defchi}
\la^m_{a+}=\bpm 0 \\ 0 \\ \chi^m_a \\ 0 \epm, \quad
\la_{m a+}=\bpm 0 \\ \chi_{m a} \\ 0 \\ 0 \epm \ .
\eea
The Majorana condition is satisfied with $(\al^m)^\ast=\al_m$, $(\beta^m)^\ast=\beta_m$ and $(\chi^m_a)^\ast=\chi_{m a}$.
Majorana conjugation acts as follows
\bea
\epbar\equiv\ep^T C=i(\epbar_m, \epbar^m), \quad
\epbar_m \equiv (\ep_m)^T C_4, \quad 
\epbar^m \equiv (\ep^m)^T C_4 \ ,
\eea
and, with our choice of the $d=4$ charge conjugation matrix,
\bea
C_4=\bpm {\bma 0 & + \\ - & 0 \ema} & \\ & {\bma 0 & - \\ + & 0 \ema} \epm \ ,
\eea
we find that
\bea
\ba[b]{rclcrcl}
\epbar^m_{-} &=& (0,0,+\al^m,0), &\quad& \epbar^m_{+} &=& (0,0,0,-\beta^m) \\[5pt]
\epbar_{m-} &=& (0,-\al_m,0,0), &\quad& \epbar_{m+} &=& (-\beta_m,0,0,0) \ .
\ea
\eea
Now we are ready to go back to the kinematical supersymmetry transformations (\ref{kinsusyA}). Defining
\bea
\label{defA}
\boxed{
A_a \equiv \frac{1}{\sqrt2}(A_{1a}+i A_{2a}), \quad
\Abar_a \equiv \frac{1}{\sqrt2}(A_{1a}-i A_{2a})
}
\eea
and the corresponding gamma matrices
\bea
\label{gaovga}
\ga &\equiv& \frac{1}{\sqrt2}(\ga_1+i\ga_2)
=\sqrt2\bpm & & & + \\ & & 0\!\! & \\ & -\!\! & & \\ 0\!\! & & & \epm \nn\\
\ov\ga &\equiv& \frac{1}{\sqrt2}(\ga_1-i\ga_2)
=\sqrt2\bpm & & & 0 \\ & & +\!\! & \\ & 0\!\! & & \\ -\!\! & & & \epm \ ,
\eea
we find that (\ref{kinsusyA}) implies
\bea
\label{Aks}
\da^\rr_{\ep_{-}}A_a &=& -i(\epbar_{m-}\ga\la_{a+}^m+\epbar^m_{-}\ga\la_{m a+})=i\sqrt2\al^m\chi_{m a} \nn\\
\da^\rr_{\ep_{-}}\Abar_a &=& -i(\epbar_{m-}\ov\ga\la_{a+}^m+\epbar^m_{-}\ov\ga\la_{m a+})=i\sqrt2\al_m\chi^m_a \ .
\eea
On another hand, from (\ref{kinsusyA}) we also deduce that
\bea
\da^\rr_{\ep_{-}}A_{(\wh I+3) a}
&=& -i(\Si^{\wh I m n}\epbar_{m-}\ga_5\la_{n a+}+\Si_{\wh I m n}\epbar^m_{-}\ga_5\la^n_{a+}) \nn\\
&=& -i(\Si^{\wh I m n}\al_m\chi_n+\Si_{\wh I m n}\al^m\chi^n) \ .
\eea
Defining
\bea
\label{defC}
\boxed{
C_{m n a} \equiv \frac{1}{\sqrt2}\Si_{\wh I m n}A_{(\wh I+3)a}, \quad
C^{m n}_a \equiv \frac{1}{\sqrt2}\Si^{\wh I m n}A_{(\wh I+3)a}
}
\eea
and using the following contraction properties of $\Si$'s (see \ref{sec-gm6})
\bea
\Si_{\wh I m n}\Si^{\wh I k l}=2(\da_m^k\da_n^l-\da_m^l\da_n^k), \quad
\Si_{\wh I m n}\Si_{\wh I k l}=2\eps_{m n k l} \ ,
\eea
we find that
\bea
\label{Cks}
\da^\rr_{\ep_{-}}C_{m n a} &=& -i\sqrt2(\al_m\chi_n-\al_n\chi_m+\eps_{m n k l}\al^k\chi^l) \nn\\
\da^\rr_{\ep_{-}}C^{m n}_a &=& -i\sqrt2(\al^m\chi^n-\al^n\chi^m+\eps^{m n k l}\al_k\chi_l) \ .
\eea
Setting $\sqrt2\al_m=\zeta_m$ (then $\sqrt2\al^m=\zeta^m$) and comparing (\ref{Aks}) and (\ref{Cks}) with (\ref{ACkinsusy}), we see that the definitions of $A_a$, $C_{m n a}$ and $\chi_{m a}$ in (\ref{defA}), (\ref{defC}) and (\ref{defchi}), respectively, provide the required embedding of the independent field components into the superfield $\phi_a$. We note also that the following property of $\Si$'s
\bea
(\Si_{\wh I m n})^\ast=\Si^{\wh I m n}=\half\eps^{m n k l}\Si_{\wh I k l}
\eea
implies the corresponding property of $C$'s.

\subsection{Dynamical supersymmetry transformation of $A_a$}
\label{sec-step5a}

Now we turn to the dynamical supersymmetry transformation of $A_{I a}$ in (\ref{dynsusyA}) and concentrate on its part corresponding to $A_a$ defined in (\ref{defA}):
\bea
\da^\rr_{\ep_{+}}A_a=\frac{i}{2\p^{+}}(\epbar_{+}\Ga_{-}\Ga^I D_I\Ga\la_{a+})
+f^{b c}{}_a\frac{1}{\p^{+}}(\p^{+}A_b\cdot\om_c) \ ,
\eea
where $\om_a$ is given in (\ref{omcomp}), $\Ga=i\ga\otimes I_8$, and
\bea
\Ga_{-}=i\ga_{-}\otimes I_8, \quad \ga_{-}=\frac{1}{\sqrt2}(\ga_0-\ga_3)
=\sqrt2\bpm & {\bma - & 0 \\ 0 & 0 \ema} \\ {\bma 0 & 0 \\ 0 & - \ema} & \epm \ .
\eea
Splitting the $SO(8)$ index $I$ into $s=1,2$ and $I^\rr=\wh I+3$, we have
\bea
\Ga^I D_I=\Ga^s D_s+\Ga^{I^\rr}D_{I^\rr}, \quad 
D_s\la_a &=& \p_s\la_a+f^{b c}{}_a A_{s b}\la_{c} \nn\\
D_{I^\rr}\la_a &=& \p_{I^\rr}\la_a+f^{b c}{}_a A_{I^\rr b}\la_c \ .
\eea
Noting that
\bea
\Ga^s D_s=i\ga^s D_s\otimes I_8, \quad 
\ga^s D_s=\sqrt2\bpm & & & \Dbar \\ & & \;D & \\ & -\Dbar\!\! & & \\ -D\!\! & & & \epm \ ,
\eea
where we defined
\bea
D=\frac{1}{\sqrt2}(D_1+i D_2), \quad \Dbar=\frac{1}{\sqrt2}(D_1-i D_2) \ ,
\eea
we easily calculate the required ingredients
\bea
\epbar_{+}\Ga_{-}\Ga^s D_s\Ga\la_{a+}
&=& \epbar_{m+}\ga_{-}\ga^s D_s\ga\la_{+a}^m
+\epbar^m_{+}\ga_{-}\ga^s D_s\ga\la_{m a+} \nn\\
&=& -2\sqrt2\beta^m D\chi_{m a}
\nn\\
\epbar_{+}\Ga_{-}\Ga^{\wh I+3}\Ga\la_{a+}
&=& \Si^{\wh I m n}\epbar_{m+}\ga_{-}\ga_5\ga\la_{n a+}
+\Si_{\wh I m n}\epbar^m_{+}\ga_{-}\ga_5\ga\la^n_{a+} \nn\\
&=& -2\Si^{\wh I m n}\beta_m\chi_{n a} \ .
\eea
Note that $\ga\la^m_{a+}=0$, as follows from (\ref{defchi}) and (\ref{gaovga}). For $\om_a$ in (\ref{omcomp}) we find
\bea
\om_a=\frac{i}{\p^{+}}(\epbar_{+}\Ga_{-}\la_{a+})
&=& -\frac{i}{\p^{+}}(\epbar_{m+}\ga_{-}\la_{a+}^m+\epbar^m_{+}\ga_{-}\la_{m a+}) \nn\\
&=& -\frac{i\sqrt2}{\p^{+}}(\beta_m\chi^m_a+\beta^m\chi_{m a}) \ .
\eea
Combining the ingredients, we obtain
\bea
\da^\rr_{\ep_{+}}A_a &=& \frac{i}{2\p^{+}}
(-2\sqrt2\beta^m D\chi_{m a}-2\Si^{\wh I m n}\beta_m D_{\wh I+3}\chi_{n a}) \nn\\
&-& i\sqrt2 f^{b c}{}_a\frac{1}{\p^{+}}\Big(\p^{+}A_{b}\cdot\frac{1}{\p^{+}}
(\beta_m\chi^m_c+\beta^m\chi_{m c})\Big) \ .
\eea
Defining now
\bea
\label{trder1}
\boxed{
\p \equiv \frac{1}{\sqrt2}(\p_1+i\p_2), \quad
\pbar \equiv \frac{1}{\sqrt2}(\p_1-i\p_2)
}
\eea
and
\bea
\label{trder2}
\boxed{
\p_{m n} \equiv \frac{1}{\sqrt2}\Si_{\wh I m n}\p_{\wh I+3}, \quad
\p^{m n} \equiv \frac{1}{\sqrt2}\Si^{\wh I m n}\p_{\wh I+3} \ ,
}
\eea
which yields an $SU(4)$ representation for the transverse $SO(8)$ derivative $\p_I$, and using the definitions (\ref{defA}) and (\ref{defC}) (which similarly represent $A_{I a}$ in terms of $A_a$ and $C_{m n a}$), we find
\bea
\label{Ads}
\boxed{
\ba[b]{rcl}
\da^\rr_{\ep_{+}}A_a &=&\dst -i\sqrt2\beta^m\frac{1}{\p^{+}}
(D\chi_{m a}+f^{b c}{}_a\p^{+}A_b\cdot\frac{1}{\p^{+}}\chi_{m c}) \\[5pt]
&&\dst -i\sqrt2\beta_m\frac{1}{\p^{+}}\Big(
D^{m n}\chi_{n a}+f^{b c}{}_a\p^{+}A_b\cdot\frac{1}{\p^{+}}\chi^m_c\Big) \ ,
\ea
}
\eea
where 
\bea
D\chi_{m a}=\p\chi_{m a}+f^{b c}{}_a A_b\chi_{m c}, \quad
D^{m n}\chi_{k a}=\p^{m n}\chi_{k a}+f^{b c}{}_a C^{m n}_b\chi_{k c} \ .
\eea
This is the result we were aiming at. Now the task is to lift this transformation to that of the LC superfield $\phi_a$.

\subsection{Dynamical supersymmetry for the LC superfield}
\label{sec-step5b}

The result (\ref{Ads}) for the dynamical supersymmetry transformation of $A_a$ can be equivalently stated as
\bea
\label{AdsQ}
\da_{\xi\ov Q} A_a &=& i\xi^m\frac{1}{\p^{+}}
\Big(\p\chi_{m a}+f^{b c}{}_a\p^{+}(A_b\frac{1}{\p^{+}}\chi_{m c})\Big) \nn\\
\da_{\xibar Q} A_a &=& i\xi_m \frac{1}{\p^{+}}
\Big(\p^{m n}\chi_{n a}+f^{b c}{}_a(C^{m n}_b\chi_{n c}+\p^{+}A_b\cdot\frac{1}{\p^{+}}\chi^m_c)\Big) \ ,
\eea
where we rescaled the parameter by defining $\xi^m=-\sqrt2\beta^m$. 

Using the relations (\ref{projphi}) and (\ref{projphibar}), we write the $O(f^0)$ part of the transformations as follows
\bea
\da_{\xi\ov Q}^{(0)}\phi_a{}_|=\xi^m d_m\frac{\p}{\p^{+}}\phi_a{}_|, \quad
\da_{\xibar Q}^{(0)}\phi_a{}_|=\xi_m d_n\frac{\p^{m n}}{\p^{+}}\phi_a{}_| \ .
\eea
As the variation of a chiral superfield must itself be chiral, we guess that~\footnote{
As $d_m\phi_|=q_m\phi_|$, there is ambiguity in lifting the projected transformation to the full superfield transformation. For the part of the transformation linear in $\phi$, chirality requires us to employ $q$'s and not $d$'s. For the nonlinear part, we will find that chirality can be achieved using $q$'s \emph{or} $d$'s. However, the two choices turn out to be \emph{identical}! This becomes manifest when using the ``coherent state operators'' $E$'s (exponents in $q/\p^{+}$ and $d/\p^{+}$): the transformations involve each $E$ with the matching $E^{-1}$, so that $q$'s can be replaced by $d$'s and vice versa \cite{BBKR}. See also Section \ref{sec-verBLG}.
}
\bea
\label{guess0}
\boxed{
\da_{\xi\ov Q}^{(0)}\phi_a=\xi^m\frac{1}{\p^{+}}(q_m\p+a_0 q^n\p_{m n})\phi_a \ ,
}
\eea
where $a_0$ is a constant to be fixed. (Note that $q^m\phi_a{}_|=d^m\phi_a{}_|=0$.) Complex conjugating this expression (see \ref{sec-cc}) and using the ``inside-out'' constraint (\ref{inout}), we find
\bea
\da_{\xibar Q}^{(0)}\phi_a=-\xi_m\frac{1}{\p^{+}}(q^m\pbar+a_0^\ast q_n\p^{m n})\phi_a \ .
\eea
The $O(f^0)$ part of (\ref{AdsQ}) is then reproduced provided we take $a_0=-1$. Using that
\bea
[q_m,q^{n k}]=\{q_m,q^n\}q^k-q^n\{q_m,q^k\}=i\sqrt2\p^{+}(\da_m^n q^k-\da_m^k q^n) \ ,
\eea
as follows from (\ref{qmqn}), we find that (\ref{guess0}) with $a_0=-1$ is the same as~\footnote{
Our equation (\ref{guess0}) agrees with equation (3.18) in \cite{ABR1}. The form (\ref{guess0a}) is given to draw an analogy with the statement in \cite{ABR1} that the lifting of $d=4$ $N=4$ results to $d=10$ $N=1$ is done by replacing $\p$ with $\nabla=\p+\frac{i}{4\sqrt2\p^{+}} d^{m n}\p_{m n}$. Although this replacement clearly does not work on the level of supersymmetry transformations, we have verified that it does work for the $O(f^1)$ part of the Lagrangian, in agreement with equation (3.32) in \cite{ABR1}. Further details will be presented elsewhere. 
}
\bea
\label{guess0a}
\boxed{
\da_{\xi\ov Q}^{(0)}\phi_a=\xi^m\frac{1}{\p^{+}}\Big(q_m\p
+\frac{i}{2\sqrt2}[q_m,q^{n k}]\frac{\p_{n k}}{\p^{+}}\Big)\phi_a \ .
}
\eea

The $O(f^1)$ part of $\da_{\xi\ov Q} A_a$ in (\ref{AdsQ}) gives
\bea
\da_{\xi\ov Q}^{(1)}\phi_a{}_| =\xi^m\frac{1}{\p^{+}}(\p^{+}\phi_b\cdot d_m\phi_c)_|f^{b c}{}_a \ .
\eea
Requiring $\da_{\xi\ov Q}^{(1)}\phi_a$ to be chiral, $d^m\da_{\xi\ov Q}^{(1)}\phi_a=0$, we are led to the following guess
\bea
\label{guess}
\boxed{
\da_{\xi\ov Q}^{(1)}\phi_a =\xi^m\frac{1}{\p^{+}}(\p^{+}\phi_b\cdot q_m\phi_c)f^{b c}{}_a \ .
}
\eea
(Thanks to $q_m=d_m-i\sqrt2\ta_m\p^{+}$ and $[b c]$ antisymmetry of $f^{b c}{}_a$, one can equivalently replace $q_m$ with $d_m$.) Next, we will verify (\ref{guess}) by showing that it reproduces the $O(f^1)$ part of $\da_{\xibar Q} A_a$ in (\ref{AdsQ}).

\subsection{Verifying the guess}
\label{sec-step6}

Complex conjugation of the $O(f^1)$ part of $\da_{\xibar Q}A_a$ in (\ref{AdsQ}) gives
\bea
\label{Abards}
\da_{\xi\ov Q}^{(1)}\Abar_a=i\xi^m f^{b c}{}_a\frac{1}{\p^{+}}
\Big(C_{m n b}\chi^n_c+\p^{+}\Abar_b\cdot\frac{1}{\p^{+}}\chi_{m c}\Big) \ .
\eea
To confirm correctness of (\ref{guess}), we should reproduce this equation by projection using
\bea
\da_{\xi\ov Q}\Abar_a=\frac{d_{[4]}}{2\p^{+}}\da_{\xi\ov Q}\phi_a{}_|, \quad 
d_{[4]}\equiv\frac{1}{4!}\eps^{i j k l}d_{i j k l} \ ,
\eea
as follows from (\ref{projphi}). First, we find that
\bea
d_{[4]}(q_m\phi_a)_|=\frac{1}{4!}\eps^{i j k l}d_{m i j k l}\phi_a{}_|=0 \ ,
\eea
as $m=1,2,3,4$ and so $d_{m i j k l}$ necessarily includes two identical $d$'s. A longer but straightforward calculation gives 
\bea
f^{b c}{}_a d_{[4]}(\p^{+}\phi_b\cdot q_m\phi_c)_|
&=& f^{b c}{}_a\frac{1}{4!}\eps^{i j k l}\Big(
\p^{+}d_{i j k l}\phi_b\cdot d_m\phi_c
+4\p^{+}d_{i j k}\phi_b\cdot d_{l m}\phi_c \nn\\
&&\hspace{20pt} 
+6\p^{+}d_{i j}\phi_b\cdot d_{k l m}\phi_c
+4\p^{+}d_i\phi_b\cdot d_{j k l m}\phi_c \Big){}_| \nn\\
&=& 2i f^{b c}{}_a\p^{+}\Big(\p^{+}\Abar_b\cdot\frac{1}{\p^{+}}\chi_{m c}+C_{m n b}\chi^n_c\Big) \ .
\eea
We then conclude that (\ref{guess}) does reproduce (\ref{Abards}). Therefore, we have shown that
\bea
\label{SYMdsLC}
\boxed{
\da_{\xi\ov Q}\phi_a=\xi^m\frac{1}{\p^{+}}\Big(\p q_m\phi_a-\p_{m n}q^n\phi_a
+f^{b c}{}_a(\p^{+}\phi_b\cdot q_m\phi_c)\Big)
}
\eea
is the correct expression for the dynamical supersymmetry transformation in the $d=10$ $N=1$ SYM. This completes the application of the algorithm of Section~\ref{sec-alg}. 

\subsection{A comment on residual gauge invariance}

When we imposed the LC gauge (\ref{LCg}), we noted that there is still residual gauge invariance with $\om_a$ satisfying $\p^{+}\om_a=0$.~\footnote{
Our choice of the solution for $A_{+a}$ in (\ref{Aplus}) fixed the gauge freedom further by imposing $\p^{-}\om_a=0$. See Section \ref{sec-cgi} for more details.
} 
The covariant derivatives $D$ and $D^{m n}$ in (\ref{Ads}) include the surviving components $A_a$ and $C^{m n}_a$ of the gauge field $A_{M a}$ in a way consistent with this residual gauge symmetry. The appearance of an explicit $A_a$, in general, would signal a breakdown of gauge invariance, but in (\ref{Ads}) we find it hit by $\p^{+}$, and $\p^{+}A_a$ \emph{is} gauge invariant under transformations with $\p^{+}\om_a=0$. One of the two $\p^{+}A_a$ in (\ref{Ads}) can be absorbed into $D$ using that
\bea
D\chi_{m a}+f^{b c}{}_a\p^{+}A_b\cdot\frac{1}{\p^{+}}\chi_{m c}
&=& \p\chi_{m a}+f^{b c}{}_a\p^{+}(A_b\frac{1}{\p^{+}}\chi_{m c}) \nn\\
&=& \p^{+}D\frac{1}{\p^{+}}\chi_{m a} \ ,
\eea
but the other $\p^{+}A_a$ remains explicit.

One could expect that the (modified) supersymmetry transformations in the LC gauge would close onto translations plus equations of motion \emph{plus residual gauge transformations}. However, the latter do not arise in the algebra. This can be explained by the fact that the constraint $\p^{+}\om_a=0$ makes the residual gauge transformations ``lower dimensional'' and such transformations cannot possibly arise in the commutator of two supersymmetry transformations. (Being ``lower dimensional,'' the residual gauge invariance also does not affect the counting of on-shell degrees of freedom.)

\subsection{SYM theories in $d<10$}

The $d=10$ $N=1$ SYM theory is the mother of all SYM theories in dimensions $d<10$ \cite{BSS,GSO}. Only one of these descendants can possibly be \emph{conformally} invariant, as the YM kinetic term, $F^2$, has (mass) dimension 4. The $d=4$ descendant is, indeed, (super)conformal: it is the $d=4$ $N=4$ SYM which is invariant under superconformal group $PSU(2,2|4)$ \cite{Nahm}. The dimensional reduction is very simple in the (on-shell) LC superspace, as the field content does not change (it is always described by the superfield $\phi_a$). All one has to do is to assume $\phi_a$ to be independent of a certain number of coordinates (and set the corresponding transverse derivatives to zero). In particular, the reduction to the $d=4$ $N=4$ SYM follows from simply setting $\p_{m n}=0$. From (\ref{SYMdsLC}) we then deduce that
\bea
\label{d4SYMdsLC}
\da_{\xi\ov Q}\phi_a=\xi^m\frac{1}{\p^{+}}\Big(\p q_m\phi_a
+f^{b c}{}_a(\p^{+}\phi_b\cdot q_m\phi_c)\Big) \ ,
\eea
which reproduces the central result of \cite{ABKR} where the nonlinear realization of the whole $PSU(2,2|4)$ was used to derive this transformation.

Note that the dimensional reduction down to $d=3$ yields the $d=3$ $N=8$ SYM, and the dynamical supersymmetry there is still given by (\ref{d4SYMdsLC}) provided we set $\p=\pbar$. This theory is not conformally invariant as one needs to introduce a mass parameter to balance the dimension of the YM kinetic term with the dimension of spacetime. As was noted by Schwarz in \cite{JS}, the (maximally supersymmetric) superconformal theory in $d=3$ must be of Chern-Simons type. This theory and its LC superfield formulation are discussed in the following sections.

\newpage
\section{The $d=3$ $N=8$ BLG theory}
\label{sec-BLG}

Maximally supersymmetric (with 16 supercharges) gauge theory in $d=3$ was discovered by Bagger, Lambert and Gustavsson \cite{BL1,Gus1,BL2}. We will introduce it here in a manner similar to that of Section \ref{sec-SYM}, following closely the presentation in \cite{BL2}. We start with a 3-Lie algebra with generators $T^a$ satisfying
\bea
\label{triple}
[T^b,T^c,T^d]=f^{b c d}{}_a T^a, \quad f^{b c d}{}_a=f^{[b c d]}{}_a \ .
\eea
The triple bracket is totally antisymmetric (it generalizes the commutator in the usual Lie algebra) which makes the structure constants $f^{b c d}{}_a$ totally antisymmetric in the three upper indices. With this 3-Lie algebra one associates gauge transformations with parameters $\wt\La^b{}_a$ such that a covariant object $X_a$ transforms as follows
\bea
\da_\La X_a=X_b\wt\La^b{}_a \ .
\eea
In order to turn its spacetime derivative $\p_\mu X_a$, $\mu=0,1,2$, into a covariant object, with $\wt\La^b{}_a$ being local ($\p_\mu\wt\La^b{}_a\neq0$), we introduce a gauge field $\wt A_\mu{}^b{}_a$ that transforms as follows
\bea
\label{AgtrBLG}
\da_\La\wt A_\mu{}^b{}_a=\p_\mu\wt\La^b{}_a-\wt\La^b{}_c\wt A_\mu{}^c{}_a+\wt A_\mu{}^b{}_c\wt\La^c{}_a
\equiv D_\mu \wt\La^b{}_a \ .
\eea
The covariant derivative is then defined as
\bea
D_\mu X_a=\p_\mu X_a-X_b\wt A_\mu{}^b{}_a
\eea
and the commutator of two covariant derivatives defines the field-strength $\wt F_{\mu\nu}{}^b{}_a$,
\bea
[D_\mu,D_\nu]X_a=X_b \wt F_{\mu\nu}{}^b{}_a, \quad
\wt F_{\mu\nu}{}^b{}_a=\p_\nu\wt A_\mu{}^b{}_a+\wt A_\nu{}^b{}_c\wt A_\mu{}^c{}_a-(\mu\lra\nu) \ ,
\eea
which also transforms covariantly.~\footnote{
Unlike the discussion of SYM in Section \ref{sec-SYM}, we find that here one \emph{does not} need any Jacobi-like identity to prove that $D_\mu X_a$ and $\wt F_{\mu\nu}{}^b{}_a$ transform covariantly. We will explain why this is the case in \ref{sec-tilde}.
}

In $d=3$, the gauge field $A_\mu$ can have either one on-shell degree of freedom (if it comes with the Yang-Mills type kinetic term) or zero on-shell degrees of freedom (if it comes with the Chern-Simons type kinetic term). The BLG theory exploits the second possibility. The bosonic degrees of freedom are now carried by 8 scalars $X^I_a$, ($I=3,\dots,10$), which realize naturally the $SO(8)$ $R$-symmetry of the $OSp(2,2|8)$ supergroup. The matching 8 fermionic degrees of freedom are carried by a 32-component Majorana spinor $\Psi_a$ satisfying an additional ``Weyl-like'' condition. The gauge field $\wt A_\mu{}^b{}_a$, carrying no on-shell degrees of freedom, however, is still required for the covariant formulation of the theory. The supersymmetry transformations are \cite{BL2}
\bea
\label{BLGsusy}
\da_\ep X_a^I &=& i\epbar\Ga^I\Psi_a \nn\\[5pt]
\da_\ep \Psi_a &=& \Ga^\mu\Ga^I\ep D_\mu X^I_a
-\frac{1}{6}X^I_b X^J_c X^K_d f^{b c d}{}_a\Ga^{I J K}\ep \nn\\[5pt]
\da_\ep \wt A_\mu{}^b{}_a &=& i\epbar\Ga_\mu\Ga^I X^I_c\Psi_d f^{c d b}{}_a \ ,
\eea
where $\ep$ is the supersymmetry parameter. The $32\times32$ matrices $(\Ga^\mu,\Ga^I)$, with $\mu=0,1,2$ and $I=3,\dots,10$, form a representation of the $d=11$ Clifford algebra which in the split form is 
\bea
\{\Ga^\mu,\Ga^\nu\}=2\eta^{\mu\nu}I_{32}, \quad
\{\Ga^\mu,\Ga^I\}=0, \quad
\{\Ga^I,\Ga^J\}=2\da^{I J}I_{32} \ ,
\eea
with $\eta^{\mu\nu}=(-++)$. The bar on the spinors denotes Majorana conjugation
\bea
\epbar\equiv\ep^T C, \quad C^T=-C, \quad C\Ga^\mu C^{-1}=-(\Ga^\mu)^T, \quad C\Ga^I C^{-1}=-(\Ga^I)^T \ .
\eea
Both $\ep$ and $\Psi_a$ are 32-component spinors satisfying the Majorana condition
\bea
\label{newM}
\ep^T C=\ep^\dagger\Ga_0, \quad \Psi_a^T C=\Psi_a^\dagger\Ga_0 \ .
\eea
In addition, they satisfy a ``Weyl-like'' condition that we state as follows~\footnote{
Choosing $\Gat=\pm\Ga^{012}=\mp\Ga_{012}$ is equivalent to choosing $\ep^{012}=-\ep_{012}=\pm1$. For this section, we do not need to specify the choice.
}
\bea
\label{newW}
\Gat\ep=\ep, \quad \Gat\Psi_a=-\Psi_a; \quad
\Ga^{\mu\nu\la}=\eps^{\mu\nu\la}\Gat \ .
\eea
(Note that, unlike the Weyl conditions (\ref{MW}) on $\ep$ and $\la_a$ in the SYM case, the ``Weyl-like'' conditions here are \emph{opposite} for $\ep$ and $\Psi_a$.) We have now all we need to analyze the closure of the supersymmetry transformations under commutation. After a long calculation, following \cite{BL2}, we find (see \ref{sec-Fierz} for useful identities)
\bea
[\da_{\ep_1},\da_{\ep_2}]X^I_a &=& v^\mu D_\mu X^I_a+X^I_b\wt\La^b{}_a \nn\\[5pt]
[\da_{\ep_1},\da_{\ep_2}]\Psi_a &=& v^\mu D_\mu\Psi_a+\Psi_b\wt\La^b{}_a
-\half\Big[v^\mu\Ga_\mu+\frac{i}{2}(\epbar_2\Ga^{I J}\ep_1)\Ga_{I J}\Big]E_a(\Psi) \nn\\[5pt]
[\da_{\ep_1},\da_{\ep_2}]\wt A_\mu{}^b{}_a &=& -v^\nu\wt F_{\nu\mu}{}^b{}_a+D_\mu\wt\La^b{}_a
+v^\nu E_{\nu\mu}(A)^b{}_a \nn\\[5pt]
&-&\frac{i}{3}(\epbar_2\Ga_\mu\Ga^{I J K L}\ep_1)X^I_c X^J_e X^K_f X^L_g f^{e f g}{}_d f^{c d b}{}_a \ ,
\eea
where we defined
\bea
v^\mu \equiv -2i(\epbar_2\Ga^\mu\ep_1), \quad
\wt\La^b{}_a \equiv -i(\epbar_2\Ga^{I J}\ep_1)X^I_c X^J_d f^{c d b}{}_a
\eea
and
\bea
E_a(\Psi) &\equiv& \Ga^\mu D_\mu\Psi_a+\half\Ga^{I J}\Psi_b X^I_c X^J_d f^{b c d}{}_a \nn\\
E_{\mu\nu}(A)^b{}_a &\equiv& \wt F_{\mu\nu}{}^b{}_a
+\eps_{\mu\nu\la}\Big(X^I_c D^\la X^J_d+\frac{i}{2}\Psibar_c\Ga^\la\Psi_d\Big)f^{c d b}{}_a \ .
\eea
Taking $E_a(\Psi)=0$ and $E_{\mu\nu}(A)^b{}_a=0$ to be the equations of motion, and imposing the following Jacobi-like constraint (the ``Fundamental Identity'') on the structure constants \cite{BL2,GNP}
\bea
\label{FI}
f^{[e f g}{}_d f^{c] d b}{}_a=0 \ ,
\eea
we conclude that the supersymmetry algebra does close (on-shell) into the translation and gauge transformation
\bea
[\da_{\ep_1},\da_{\ep_2}]X^I_a &=& v^\nu\p_\nu X^I_a
+X^I_b(\wt\La^b{}_a-v^\nu\wt A_\nu{}^b{}_a) \nn\\[5pt]
[\da_{\ep_1},\da_{\ep_2}]\Psi_a &=& v^\nu\p_\nu\Psi_a
+\Psi_b(\wt\La^b{}_a-v^\nu\wt A_\nu{}^b{}_a) \nn\\[5pt]
[\da_{\ep_1},\da_{\ep_2}]\wt A_\mu{}^b{}_a &=& v^\nu\p_\nu\wt A_\mu{}^b{}_a
+D_\mu(\wt\La^b{}_a-v^\nu\wt A_\nu{}^b{}_a) \ .
\eea
To derive the required equation of motion for the scalars, we apply supersymmetry variation to $E_a(\Psi)$ and, after using the Fundamental Identity (\ref{FI}), we find that
\bea
\label{varEpsi}
\da_\ep E_a(\Psi)=\Ga^I\ep E_a^I(X)+\half\Ga^I\Ga^{\mu\nu}\ep X_b^I E_{\mu\nu}(A)^b{}_a \ ,
\eea
where we defined
\bea
E_a^I(X) &\equiv& D^\mu D_\mu X^I_a-\frac{i}{2}\Psibar_c\Ga^{I J}X^J_d\Psi_b f^{c d b}{}_a \nn\\[3pt]
&+& \half f^{b c d}{}_a f^{f g e}{}_d(X^J_b X^J_f)(X^K_c X^K_g)X^I_e \ .
\eea
This implies that $E_a^I(X)=0$ is the third equation of motion, and the BLG theory is now fully defined.~\footnote{
We remind that in this paper we intentionally do not introduce the metric for gauge indices which is needed for the Lagrangian formulation. 
}

\section{Dynamical supersymmetry in $d=3$ $N=8$ BLG theory}
\label{sec-dsBLG}

In this section, we will apply the algorithm of Section \ref{sec-alg} to the BLG theory described above. The analysis is quite similar to that in the SYM case, but there are some important differences: 1) $\ep$ and $\Psi_a$ have \emph{opposite} chiralities with respect to $\Gat$ and 2) all three components of the gauge field $\wt A_\mu{}^b{}_a$ become dependent in the LC gauge. Still, the independent components will fit into the LC superfield $\phi_a$ and our algorithm will give its dynamical supersymmetry variation.

\subsection{LC coordinates and LC gauge}

As we defined $\mu=0,1,2$, we have to modify the definition (\ref{LCx}) of LC coordinates. We take
\bea
x^{\pm}=\frac{1}{\sqrt2}(x^0\pm x^1) \ .
\eea
The indices $\mu=0,1,2$ and $I=3,4,5,6,7,8,9,10$ will be split as follows
\bea
\label{muI}
\mu=(+,-,2), \quad I=(s,I^\rr), \quad s=3,4, \quad I^\rr=\wh I+4 \ ,
\eea
with $\wh I=1,2,3,4,5,6$ as before.
Accordingly, we will take the following representation for gamma matrices
\bea
\label{ga0134}
\Ga^0=i\ga^0\otimes I_8, \quad \Ga^1=i\ga^3\otimes I_8; \quad
\Ga^3=i\ga^1\otimes I_8, \quad \Ga^4=i\ga^2\otimes I_8 \ ,
\eea
and
\bea
\Ga^2 &=& -\Ga_\ast, \quad
\Ga_\ast=\ga_5\otimes\bpm I_4 & 0 \\ 0 & -I_4 \epm \nn\\[5pt]
\Ga^{I^\rr} &=& i\ga_5\otimes\wh\Ga^{I^\rr-4}, \quad
\wh\Ga^{\wh I}=\bpm 0 & \Si^{\wh I m n} \\ \Si_{\wh I m n} & 0 \epm \ .
\eea
Our projectors $P_{\pm}$ then remain the same as in Sections \ref{sec-step1} and \ref{sec-step4}. We will take $\eps_{012}=+1$ so that $\Gat=\Ga_{012}$ and the ``Weyl-like'' conditions (\ref{newW}) become
\bea
\Ga_{012}\ep=\ep, \quad \Ga_{012}\Psi_a=-\Psi_a \ .
\eea
Noting that
\bea
P_{+} &=& -\half\Ga^{-}\Ga^{+}=-\qter(\Ga^0-\Ga^1)(\Ga^0+\Ga^1) \nn\\
&=& \half(I_{32}-\Ga^{0}\Ga^1)
=\half(I_{32}-\eps^{012}\Ga^2\Gat) \ ,
\eea
we find that $\ep^{012}=-1$ implies  
\bea
\ep_{\pm} \equiv P_{\pm}\ep=\half(I_{32}\pm\Ga^2)\ep, \quad 
\Psi_{\pm a} \equiv P_{\pm}\Psi_a=\half(I_{32}\mp\Ga^2)\Psi_a \ .
\eea
With our choice $\Ga^2=-\Ga_\ast$, we then have
\bea
\label{pmW}
\boxed{
\ba[b]{l}
\Ga_\ast\ep_{-}=+\ep_{-}, \quad
\Ga_\ast\Psi_{a+}=+\Psi_{a+}
\\
\Ga_\ast\ep_{+}=-\ep_{+}, \quad
\Ga_\ast\Psi_{a-}=-\Psi_{a-} \ .
\ea
}
\eea
After these preparations, we finally state that our choice of the LC gauge is~\footnote{
As in the SYM case, there remains residual gauge invariance with $\wt\La^b{}_a$ satisfying $\p^{+}\wt\La^b{}_a=0$.
}
\bea
\label{newLCg}
\boxed{
\text{LC gauge: } \quad \wt A_{-}{}^b{}_a=0 \ ,
}
\eea
which implies that
\bea
\wt A^{+b}{}_a=0, \quad \wt F_{\mu -}{}^b{}_a=-\p^{+}\wt A_\mu{}^b{}_a \ .
\eea

\subsection{Dependent field components}

Let us now use the equations of motion
\bea
E_a(\Psi) &\equiv& \Ga^\mu D_\mu\Psi_a+\half\Ga^{I J}\Psi_b X^I_c X^J_d f^{b c d}{}_a=0 \nn\\
E_{\mu\nu}(A)^b{}_a &\equiv& \wt F_{\mu\nu}{}^b{}_a
+\eps_{\mu\nu\la}\Big(X^I_c D^\la X^J_d+\frac{i}{2}\Psibar_c\Ga^\la\Psi_d\Big)f^{c d b}{}_a=0  \quad
\eea
to solve for field components that become dependent in the LC gauge. (We will not need the $E_a^I(X)=0$ equation of motion; all $X_a^I$ are independent.) Hitting $E_a(\Psi)=0$ with $\Ga_{-}$ and solving the resulting equation for $\Psi_{a-}$, we find (cf. \cite{BN})
\bea
\label{psisol}
\boxed{
\Psi_{a-}=\frac{1}{2\p^{+}}\Ga_{-}(\Ga^2 D_2\Psi_{a+}+\half\Ga^{I J}\Psi_{b+} X^I_c X^J_d f^{b c d}{}_a) \ ,
}
\eea
where
\bea
\label{d2cov}
\boxed{
D_2\Psi_{a+}=\p_2\Psi_{a+}-\Psi_{b+}\wt A_2{}^b{}_a \ .
}
\eea
From $E_{+-}(A)^b{}_a=0$, using that $\eps_{012}=+1$ implies $\eps_{+-2}=-1$, we obtain
\bea
\label{A+sol}
\wt A_{+}{}^b{}_a=-\frac{1}{\p^{+}}(X^I_c D_2 X^I_d+\frac{i}{2}\Psibar_c\Ga_2\Psi_d)f^{c d b}{}_a \ ,
\eea
whereas from $E_{2-}(A)^b{}_a=0$, using that $\ep_{2-+}=+1$, we find
\bea
\label{A2sol}
\boxed{
\wt A_2{}^b{}_a=\frac{1}{\p^{+}}(X^I_c \p^{+} X^I_d-\frac{i}{2}\Psibar_{c+}\Ga_{-}\Psi_{d+})f^{c d b}{}_a \ .
}
\eea
Together with $\wt A_{-}{}^b{}_a=0$, this shows explicitly that there are no independent degrees of freedom in the gauge field $\wt A_\mu{}^b{}_a$. (Note that we will not use the solution for $\wt A_{+}{}^b{}_a$ in what follows, similar to the SYM case.)

\subsection{Modified supersymmetry transformations}

As in the SYM case, we need to modify supersymmetry transformations by adding to them compensating gauge transformations in order to preserve the LC gauge (\ref{newLCg}). The combined supersymmetry and gauge transformations in the BLG theory are
\bea
\da^\rr_\ep X_a^I &=& i\epbar\Ga^I\Psi_a
+X^I_b\wt\La^b{}_a \nn\\[5pt]
\da^\rr_\ep \Psi_a &=& \Ga^\mu\Ga^I\ep D_\mu X^I_a
-\frac{1}{6}X^I_b X^J_c X^K_d f^{b c d}{}_a\Ga^{I J K}\ep
+\Psi_b\wt\La^b{}_a \nn\\[5pt]
\da^\rr_\ep \wt A_\mu{}^b{}_a &=& i\epbar\Ga_\mu\Ga^I X^I_c\Psi_d f^{c d b}{}_a 
+D_\mu\wt\La^b{}_a \ .
\eea
Requiring that $\da^\rr_\ep \wt A_{-}{}^b{}_a=0$, we find
\bea
\label{lacomp}
\boxed{
\wt\La^b{}_a=\frac{i}{\p^{+}}(\epbar_{+}\Ga_{-}\Ga^I X^I_c\Psi_{d+})f^{c d b}{}_a \ .
}
\eea
For the following, we will only need the (modified) supersymmetry transformation of $X_a^I$. Separating the $\ep_{+}$ and $\ep_{-}$ parts, we have
\bea
\label{kinsusyX}
\da^\rr_{\ep_{-}} X_a^I = i\epbar_{-}\Ga^I\Psi_{a+}
\eea
and
\bea
\label{dynsusyX}
\da^\rr_{\ep_{+}} X_a^I = i\epbar_{+}\Ga^I\Psi_{a-}+X^I_b\wt\La^b{}_a \ ,
\eea
where we need to substitute (\ref{psisol}) for $\Psi_{a-}$ and (\ref{lacomp}) for $\wt\La^b{}_a$. As in the SYM case, we clearly see that $\ep_{-}$ transformations should be identified with the kinematical supersymmetry, whereas $\ep_{+}$ transformations with the dynamical supersymmetry.

\subsection{Identifying superfield components}

Noting that 
\begin{itemize}
\item[1)]
(\ref{kinsusyX}) is identical to (\ref{kinsusyA}), modulo replacing $A_{I a}$ with $X_a^I$ and $\la_{a+}$ with $\Psi_{a+}$;
and 
\item[2)]
the conditions (\ref{pmW}) on $\ep_{-}$ and $\Psi_{a+}$ are identical to the Weyl conditions on $\ep_{-}$ and $\la_{a+}$ (and they are all Majorana spinors, with the same charge conjugation matrix $C$);
\end{itemize}
we can copy the corresponding results from Section \ref{sec-step4}. We therefore write
\bea
\ep_{-}=\bpm \ep^m_{-} \\ \ep_{m-} \epm, \quad
\Psi_{a+}=\bpm \psi^m_{a+} \\ \psi_{m a-} \epm ,
\eea
where
\bea
\ep^m_{-}=\bpm 0 \\ 0 \\ 0 \\ \al^m \epm, \;
\ep_{m-}=\bpm -\al_m \\ 0 \\ 0 \\ 0 \epm; \quad
\psi^m_{a+}=\bpm 0 \\ 0 \\ \chi^m_a \\ 0 \epm, \;
\psi_{m a+}=\bpm 0 \\ \chi_{m a} \\ 0 \\ 0 \epm ,\ 
\eea
and furthermore
\bea
\label{defAC}
&& A_a=\frac{1}{\sqrt2}(X_a^3+i X_a^4), \quad
\Abar_a=\frac{1}{\sqrt2}(X_a^3-i X_a^4) \nn\\
&& C_{m n a}=\frac{1}{\sqrt2}\Si_{\wh I m n}X_a^{\wh I+4}, \quad
C^{m n}_a=\frac{1}{\sqrt2}\Si^{\wh I m n}X_a^{\wh I+4} \ .
\eea
These definitions provide the correct embedding of the independent fields in the BLG theory into the LC superfield $\phi_a$ (cf. \cite{BN}).

We are now in position to rewrite the solution for $\wt A_2{}^b{}_a$, as given in (\ref{A2sol}), in terms of $A_a$, $C_{m n a}$ and $\chi_{m a}$. For the bosonic part, we find 
\bea
X^{I}_c\p^{+}X^{I}_d=A_c\p^{+}\Abar_d+\Abar_c\p^{+}A_d+\half C_{m n c}\p^{+}C^{m n}_d \ ,
\eea
where we used that $\Si_{I m n}\Si^{J m n}=4\da_I^J$. For the fermionic bilinear, we obtain
\bea
\Psibar_{c+}\Ga_{-}\Psi_{d+}
&=& -(\psibar_{m c+}\ga_{-}\psi^m_{d+}+\psibar^m_{c+}\ga_{-}\psi_{m d+}) \nn\\
&=& -\sqrt2(\chi_{m c}\chi^m_d+\chi^m_c\chi_{m d}) \ .
\eea
Combining the two expressions, we rewrite (\ref{A2sol}) as
\bea
\label{A2sol2}
\wt A_2{}^b{}_a=\frac{1}{\p^{+}}\Big(A_c\p^{+}\Abar_d+\Abar_c\p^{+}A_d+\half C_{m n c}\p^{+}C^{m n}_d
+i\sqrt2\chi^m_c\chi_{m d}\Big)f^{b c d}{}_a \ . \quad
\eea
This result will be used shortly.

\subsection{Dynamical supersymmetry transformation of $A_a$}

We will now proceed to find the expression for the dynamical supersymmetry transformation of the lowest component of the LC superfield $\phi_a$. From (\ref{dynsusyX}) and the definition of $A_a$ in (\ref{defAC}), we have
\bea
\da^\rr_{\ep_{+}}A_a=i\epbar_{+}\Ga\Psi_{a-}+A_b\wt\La^b{}_a, \quad
\Ga=i\ga\otimes I_8, \quad
\ga=\frac{1}{\sqrt2}(\ga^1+i\ga^2) \ .
\eea
Substituting (\ref{psisol}) and (\ref{lacomp}), we obtain
\bea
\label{BLGdsA}
\da_{\ep_{+}}^\rr A_a &=& \frac{i}{2\p^{+}}(\epbar_{+}\Ga_{-}\Ga D_2\Psi_{a+})
-\frac{i}{4\p^{+}}(\epbar_{+}\Ga_{-}\Ga\Ga^{I}X^{I}_c\Ga^{J}X^{J}_d\Psi_{b+})f^{b c d}{}_a \nn\\
&&\hspace{50pt}
+A_b\frac{i}{\p^{+}}(\epbar_{+}\Ga_{-}\Ga^{I}X^{I}_c\Psi_{d+})f^{b c d}{}_a \ ,
\eea
where we used that
\bea
&& \Ga^2\Psi_{a+}=-\Ga_\ast\Psi_{a+}=-\Psi_{a+}, \quad 
\Ga^{I}\Ga_{-}=-\Ga_{-}\Ga^{I} \nn\\[5pt]
&& \Ga^{I J}X^{I}_c X^{J}_d f^{b c d}{}_a=\Ga^{I}\Ga^{J}X^{I}_c X^{J}_d f^{b c d}{}_a \ .
\eea
To proceed, we need the decomposition of the 32-component spinor $\ep_{+}$ into 1-component spinors with the $SU(4)$ index. We note that $\ep_{+}$ satisfies different constraints as compared to the SYM case. Taking into account (\ref{pmW}), we have
\bea
\ep_{+}^T C=\ep_{+}^\dagger\Ga_0, \quad
\Ga_\ast\ep_{+}=-\ep_{+}, \quad
P_{+}\ep_{+}=\ep_{+}
\eea
(cf. $\Ga_\ast\ep_{+}=+\ep_{+}$ in the SYM case). It then follows that we should take
\bea
\ep_{+}=\bpm \ep_{+}^m \\ \ep_{m+} \epm, \quad
\ep_{+}^m=\bpm 0 \\ \eta^m \\ 0 \\ 0 \epm, \quad
\ep_{m+}=\bpm 0 \\ 0 \\ \eta_m \\ 0 \epm, \quad
(\eta^m)^\ast=\eta_m \ ,
\eea
where we called the independent components $\eta^m$ (and not $\beta^m$ as in the SYM case) to emphasize the difference. For the Majorana conjugated spinors we then have
\bea
&& \epbar_{+} = i(\epbar_{m+},\epbar^m_{+}), \quad
\Psibar_{a+} = i(\psibar_{m a+},\psibar^m_{a+}) \nn\\[5pt]
&& \epbar^m_{+}=(-\eta^m,0,0,0), \quad
\psibar^m_{a+}=(0,0,0,-\chi^m_a) \nn\\[5pt]
&& \epbar_{m+}=(0,0,0,-\eta_m), \quad
\psibar_{m a+}=(-\chi_{m a},0,0,0) \ .
\eea
Returning to (\ref{BLGdsA}), we observe that $X^I_a$ enter only via the matrix $\Ga^I X^I_a$. Splitting the $SO(8)$ index $I$ into $s=3,4$ and $I^\rr=\wh I+4$, we find that
\bea
\Ga^I X^I_a
&=& \bpm i\ga^s X^s_a\da^m_n & i\sqrt2\ga_5 C^{m n}_a \\[10pt]
i\sqrt2\ga_5 C_{m n a} & i\ga^s X^s_a\da_m^n \epm 
\nn\\[3pt]
\ga^s X^s_a
&=& \sqrt2\bpm & & & \Abar_a \\ & & \;A_a & \\ & -\Abar_a\!\! & & \\ -A_a\!\! & & & \epm \ . \qquad
\eea
Noting that $\Ga_{-}=i\ga_{-}\otimes I_8$ and $\Ga=i\ga\otimes I_8$, with $\ga_{-}$ and $\ga$ the same as in the SYM case, we find for the fermionic bilinears in (\ref{BLGdsA})
\bea
\epbar_{+}\Ga_{-}\Ga\Psi_{a+} &=& -i\Big(
\epbar_{m+}\ga_{-}\ga\psi^m_{a+}
+\epbar^m_{+}\ga_{-}\ga\psi_{m a+}\Big) 
\nn\\
\epbar_{+}\Ga_{-}\Ga^I X^I_c\Psi_{d+} &=& -i\Big(
\epbar_{m+}\ga_{-}\ga^s X^s_c\psi^m_{d+}
+\epbar^m_{+}\ga_{-}\ga^s X^s_c\psi_{m d+} \nn\\
&&+\sqrt2 C^{m n}_c\epbar_{m+}\ga_{-}\ga_5\psi_{n d+}
+\sqrt2 C_{m n c}\epbar^m_{+}\ga_{-}\ga_5\psi^n_{d+} \Big) 
\nn\\
\epbar_{+}\Ga_{-}\Ga\Ga^I X^I_c\Ga^J X^J_d\Psi_{b+} &=& i\Big(
\epbar_{m+}\ga_{-}\ga\ga^s X^s_c\ga^t X^t_d\psi^m_{b+}
+\epbar^m_{+}\ga_{-}\ga\ga^s X^s_c\ga^t X^t_d\psi_{m b+} \nn\\
&& \hspace{-60pt}
+\sqrt2 C^{m k}_d\epbar_{m+}\ga_{-}\ga\ga^s X^s_c\ga_5\psi_{k b+}
+\sqrt2 C_{m k d}\epbar^m_{+}\ga_{-}\ga\ga^s X^s_c\ga_5\psi^k_{b+} \nn\\
&& \hspace{-60pt}
-\sqrt2 C^{m k}_c\epbar_{m+}\ga_{-}\ga\ga^s X^s_d\ga_5\psi_{k b+}
-\sqrt2 C_{m k c}\epbar^m_{+}\ga_{-}\ga\ga^s X^s_d\ga_5\psi^k_{b+} \nn\\
&& \hspace{-60pt}
+2C^{m n}_c C_{n k d}\epbar_{m+}\ga_{-}\ga\psi^k_{b+}
+2 C_{m n c}C^{n k}_d\epbar^m_{+}\ga_{-}\ga\psi_{k b+} \Big) \ ,
\eea
and furthermore
\bea
\epbar_{+}\Ga_{-}\Ga\Psi_{a+} &=& 2i\eta^m\chi_{m a} 
\nn\\
\epbar_{+}\Ga_{-}\Ga^I X^I_c\Psi_{d+} &=& -2i\Big(
A_c\eta_m\chi^m_d-\Abar_c\eta^m\chi_{m d} \nn\\
&&\hspace{30pt}
+C_{m n c}\eta^m\chi^n_d-C^{m n}_c\eta_m\chi_{n d}\Big) 
\nn\\
\epbar_{+}\Ga_{-}\Ga\Ga^I X^I_c\Ga^J X^J_d\Psi_{b+} &=& 4i\Big(
A_c\Abar_d\eta^m\chi_{m b}+(C_{m k c}A_d-A_c C_{m k d})\eta^m\chi^k_b \nn\\
&&\hspace{60pt}
-C_{m n c}C^{n k}_d\eta^m\chi_{k b}\Big) \ . 
\eea
Combining these results into (\ref{BLGdsA}), using that $D_2\chi_{m a}=\p_2\chi_{m a}-\chi_{m b}\wt A_2{}^b{}_a$ with $\wt A_2{}^b{}_a$ given in (\ref{A2sol2}), and separating $\eta^m$ from $\eta_m$ transformations, we find that the dynamical supersymmetry transformations of $A_a$ are as follows
\bea
\label{dsAblg1}
\boxed{
\da_{\etabar Q}A_a=2\eta_m A_b\frac{1}{\p^{+}}(A_c\chi^m_d-C^{m n}_c\chi_{n d})f^{b c d}{}_a
}
\eea
and
\bea
\label{dsAblg2}
\da_{\eta\ov Q}A_a &=& -\eta^m\frac{\p_2}{\p^{+}}\chi_{m a} \nn\\
&&\hspace{-40pt} 
+\ \eta^m\Big\{\frac{1}{\p^{+}}\Big[\chi_{m b}\frac{1}{\p^{+}}\Big(
A_c\p^{+}\Abar_d+\Abar_c\p^{+}A_d+\half C_{n k c}\p^{+}C^{n k}_d+i\sqrt2\chi^n_c\chi_{n d}\Big)\Big] \nn\\
&&\hspace{50pt}
+\frac{1}{\p^{+}}\Big(A_c\Abar_d\chi_{m b}+2C_{m n c}A_d\chi^n_b-C_{m n c}C^{n k}_d\chi_{k b}\Big) \nn\\
&&\hspace{50pt}
+2A_b\frac{1}{\p^{+}}\Big(-\Abar_c\chi_{m d}+C_{m n c}\chi^n_d\Big) \Big\} f^{b c d}{}_a \ .
\eea
Our next task is to lift these transformations to the superfield form.

\subsection{Dynamical supersymmetry for the LC superfield}

Let us first analyze the $O(f^0)$ part of the dynamical supersymmetry transformations (\ref{dsAblg1}) and (\ref{dsAblg2}). We have
\bea
\da_{\eta\ov Q}^{(0)}A_a=-\eta^m\frac{\p_2}{\p^{+}}\chi_{m a}, \quad
\da_{\etabar Q}^{(0)}A_a=0 \ .
\eea
Using the definitions of the superfield components in (\ref{projphi}), and requiring the variation of the superfield $\phi_a$ to be chiral, we are led to the following superfield transformations
\bea
\label{dsBLGfree}
\boxed{
\da_{\eta\ov Q}^{(0)}\phi_a=i\eta^m q_m\frac{\p_2}{\p^{+}}\phi_a, \quad
\da_{\etabar Q}^{(0)}\phi_a=i\eta_m q^m\frac{\p_2}{\p^{+}}\phi_a \ .
}
\eea
Note that the $\eta_m$ transformation follows from the $\eta^m$ one by complex conjugation and the use of the ``inside-out'' constraint (\ref{inout}). It does give the correct projection as $q^m\phi_|=d^m\phi_|=0$. 

To analyze the $O(f^1)$ part of the supersymmetry transformations, we first rewrite (\ref{dsAblg1}) as
\bea
\da_{\etabar Q}^{(1)}(\p^{+}\phi_a)_|
&=& 2\eta_m(\p^{+}\phi_b)\frac{1}{\p^{+}}\Big[
(\p^{+}\phi_c)(i\p^{+}d^m\phibar_d) \nn\\
&&\hspace{80pt}
-(\frac{i}{\sqrt2}d^{m n}\phibar_c)(-i\p^{+}d_n\phi_d)\Big]f^{b c d}{}_a{}_| \ . \qquad
\eea
Omitting the projection signs, we obtain the natural guess for the full superfield transformation law 
\bea
\label{dsBLG}
\boxed{
\ba[b]{rcl}
\da_{\etabar Q}^{(1)}\phi_a &=&\dst 2i\eta_m\frac{1}{\p^{+}}(\p^{+}\phi_b\cdot\frac{1}{\p^{+}}W_{c d}^m)f^{b c d}{}_a 
\\[10pt]
W_{c d}^m &\equiv&\dst \p^{+}\phi_c\cdot\p^{+}d^m\phibar_d-\frac{i}{\sqrt2}\p^{+}d_n\phi_c\cdot d^{m n}\phibar_d \ .
\ea
}
\eea
We will check next that this form is consistent with the chirality of $\phi_a$ and that it does reproduce the $O(f^1)$ part of (\ref{dsAblg2}).

\subsection{Verifying the guess}
\label{sec-verBLG}

Using $d^k\phi=0$, $\{d^k,d_n\}=-i\sqrt2\da_n^k\p^{+}$ and $d^{k m n}\phibar=-i\sqrt2\eps^{k m n l}\p^{+}d_l\phi$, which is a consequence of (\ref{inout}), we find that
\bea
d^k W_{c d}^m=\p^{+}(\p^{+}\phi_c\cdot\p^{+}d^{k m}\phibar_d)
+\eps^{k m n l}\p^{+}d_n\phi_c\cdot\p^{+}d_l\phi_d \ .
\eea
The second term is symmetric under $(c\lra d)$ and vanishes when contracted with $f^{b c d}{}_a=f^{[b c d]}{}_a$. The first term, when substituted into (\ref{dsBLG}), yields $\p^{+}\phi_b\cdot\p^{+}\phi_c$, which is symmetric under $(b\lra c)$ and also vanishes when contracted with $f^{b c d}{}_a$. This then proves that (\ref{dsBLG}) is chiral,
\bea
d^k(\da_{\etabar Q}^{(1)}\phi_a)=0 \ .
\eea

It is also possible to transform (\ref{dsBLG}) to the form that contains $q$'s instead of $d$'s, which makes the chirality manifest. To this end, we rewrite (\ref{dsBLG}) as
\bea
\label{dsBLG1}
\da_{\etabar Q}^{(1)}\phi_a &=& \frac{\eps^{m n k l}}{3\sqrt2}\eta_m\frac{1}{\p^{+}}\Big(
\p^{+}\phi_b\cdot\frac{1}{\p^{+}}\Big[
\p^{+}\phi_c\cdot d_{n k l}\phi_d
+3\p^{+}d_n\phi_c\cdot d_{k l}\phi_d\Big]\Big)f^{b c d}{}_a \nn\\
&&\hspace{-40pt} 
=\frac{\eps^{m n k l}}{3!\sqrt2}\eta_m\frac{\p}{\p\zeta^{n k l}}
\frac{1}{\p^{+}}\Big(\p^{+}\phi_b\cdot\frac{1}{\p^{+}}\Big(
\p^{+2}E_\zeta\phi_c\cdot\p^{+2}E_{-\zeta}\phi_d\Big)\Big)_{\big|\zeta=0}f^{b c d}{}_a \ ,
\eea
where in the first line we used the ``inside-out'' constraint (\ref{inout}), and in the second line we introduced the ``coherent state operators'' \cite{ABKR,BBKR}
\bea
E_\zeta=\exp(\zeta^m d_m/\p^{+}) \ ,
\eea
and used the $[b c d]$ symmetry of $f^{b c d}{}_a$. As $d_m/\p^{+}$ differs from $q_m/\p^{+}$ by a constant, $i\sqrt2\ta_m$, the $(E_\zeta,E_{-\zeta})$ structure of (\ref{dsBLG1}) makes it obvious that all the $d$'s there can be replaced by $q$'s. The chirality of $\da_{\etabar Q}^{(1)}\phi_a$ is then manifest.

The verification that (\ref{dsBLG}) reproduces (\ref{dsAblg2}) is straightforward but tedious. The basic idea is to use that
\bea
\label{vs1}
\da_{\etabar Q}^{(1)}\Abar_a=\frac{d_{[4]}}{2\p^{+}}\Big(\da_{\etabar Q}^{(1)}\phi_a\Big){}_| \ ,
\eea
as follows from the ``inside-out'' constraint (\ref{inout}), and then conjugate the result to find
\bea
\label{vs2}
\da_{\eta\ov Q}^{(1)}A_a=\Big(\da_{\etabar Q}^{(1)}\Abar_a\Big)^\ast \ .
\eea
For this calculation, the following identities are helpful
\bea
d_l d^m\phibar &=& -i\sqrt2\da_l^m\p^{+}\phibar \nn\\
d_l d^{m n}\phibar &=& -i\sqrt2(\da_l^m d^n-\da_l^n d^m)\p^{+}\phibar \nn\\
d_{k l}d^{m n}\phibar &=& 2(\da_k^m\da_l^n-\da_k^n\da_l^m)\p^{+2}\phibar \ .
\eea
It also helps to group the terms by their field content, so that
\bea
\label{ACchi}
\da_{\etabar Q}^{(1)}\Abar_a=\eta_m\frac{1}{\p^{+2}}\Big[ 
(A A\chi)+(C C\chi)+(A C\chi)+(\chi\chi\chi)\Big] f^{b c d}{}_a \ .
\eea
Using the fact that the antisymmetrization in five $SU(4)$ indices gives zero,
\bea
(C^{[i j},C^{k l},\chi^{m]})=0 \ ,
\eea
we find the following identity
\bea
\label{CCchiID}
(C^{i j},C_{i j},\chi^m)=2(C_{k n},C^{k m},\chi^n)+2(C^{k m},C_{k n},\chi^n) \ ,
\eea
which is needed to simplify the $(C C\chi)$ terms. After a long calculation that also uses the $[b c d]$ symmetry of $f^{b c d}{}_a$, we arrive at the following result
\bea
\da_{\eta\ov Q}^{(1)}A_a &=& \eta^m\frac{1}{\p^{+}}\Big[
2\chi_{m b}\frac{1}{\p^{+}}(\Abar_c\p^{+}A_d)
-2\p^{+}A_b\cdot\frac{1}{\p^{+}}(\Abar_c\chi_{m d}) \nn\\
&&\hspace{40pt}
-2\chi_{k b}\frac{1}{\p^{+}}(C_{m n c}\p^{+}C^{n k}_d) 
-2\p^{+}A_b\cdot\frac{1}{\p^{+}}(\chi^n_c C_{m n d}) \nn\\
&&\hspace{40pt}
+i\sqrt2\chi_{m b}\frac{1}{\p^{+}}(\chi^n_c\chi_{n d}) \Big]f^{b c d}{}_a \ .
\eea
As the $O(f^1)$ part of (\ref{dsAblg2}) can also be brought to this form, this confirms correctness of (\ref{dsBLG}). Therefore, (\ref{dsBLG}) together with (\ref{dsBLGfree}) gives the dynamical supersymmetry transformation of the LC superfield $\phi_a$ in the BLG theory. This is the main result of our work.

\subsection{A comment on residual gauge invariance}
\label{sec-cgi}

When we imposed the LC gauge (\ref{newLCg}), we noted that there is still residual gauge invariance with $\wt\La^b{}_a$ satisfying $\p^{+}\wt\La^b{}_a=0$. However, our expression (\ref{dsAblg2}) is clearly not invariant under these residual gauge transformations, as $\p_2$ appears there without a gauge field that would cancel the $\p_2\wt\La^b{}_a$ part of the variation. The place where we lost the gauge invariance is in equation (\ref{A2sol}). The $E_{2-}(A)^b{}_a=0$ equation of motion, in fact, has the following solution~\footnote{
We find that $\da_\La W_{c d}f^{c d b}{}_a=(W_{c f}\wt\La^f{}_d+W_{f d}\wt\La^f{}_c)f^{c d b}{}_a$. One then has to use the Fundamental Identity in the form (\ref{FI4}) to prove that $\wt A_2{}^b{}_a$ transforms as required. See also \ref{sec-tilde}.
}
\bea
\wt A_2{}^b{}_a=\wt B_2{}^b{}_a+\frac{1}{\p^{+}}W_{c d}f^{c d b}{}_a, \quad
W_{c d}\equiv X^I_c\p^{+}X^I_d-\frac{i}{2}\Psibar_{c+}\Ga_{-}\Psi_{d+} \ ,
\eea
where 
\bea
\p^{+}\wt B_2{}^b{}_a=0, \quad
\da_\La \wt B_2{}^b{}_a=\p_2\wt\La^b{}_a-\wt\La^b{}_c\wt B_2{}^c{}_a+\wt B_2{}^b{}_c\wt\La^c{}_a \ .
\eea
We implicitly assumed that $\wt B_2{}^b{}_a=0$, which then requires $\p_2\wt\La^b{}_a=0$ for consistency. Analogously, dropping the corresponding ``integration constant'' in (\ref{A+sol}), we also imposed $\p^{-}\wt\La^b{}_a=0$. Altogether, our choice of the LC gauge (\ref{newLCg}) and the form of the solutions for dependent gauge field components (\ref{A+sol}) and (\ref{A2sol}) led to fixing the gauge freedom completely (only transformations with rigid $\wt\La^b{}_a$ are still a symmetry).

We could restore $\wt B_2{}^b{}_a$ which would then turn every $\p_2$ into a covariant derivative $D_2$. However, $\wt B_2{}^b{}_a$ would be a ``supersymmetry singlet'' (i.e. invariant under supersymmetry) and as such inessential for our construction.~\footnote{
For an interesting example of a supersymmetry singlet which acquires nonzero boundary-localized supersymmetry variation, see \cite{BvN}.
}

\section{Conclusion}

Light-cone (LC) superspace provides a convenient foundation for describing maximally supersymmetric gauge theories. A single scalar chiral superfield $\phi_a$, satisfying the additional ``inside-out'' constraint (\ref{inout}), describes all on-shell degrees of freedom. No auxiliary fields are required, in sharp distinction with the conventional superspace formulations. The key ingredient in the LC superspace formulation is the dynamical supersymmetry transformation $\da_{\ep\Qbar}\phi_a$. The conjugated transformation, $\da_{\epbar Q}\phi_a$, follows via complex conjugation and the use of the ``inside-out'' constraint. The Hamiltonian shift, $\da_{\mc{P}^{-}}\phi_a$, follows by commuting $\da_{\ep\Qbar}$ with $\da_{\epbar Q}$, and it encodes the dynamics of the theory as the equations of motion are $\p^{-}\phi_a=i\da_{\mc{P}^{-}}\phi_a$ \cite{BBKR}.

In this paper, we presented an explicit derivation of $\da_{\ep\Qbar}\phi_a$ for the cases of known maximally supersymmetric super-Yang-Mills and super-Chern-Simons theories, starting from their covariant formulations. 

In the case of $d=10$ $N=1$ SYM, our result for $\da_{\ep\Qbar}\phi_a$ is given in (\ref{SYMdsLC}). Maximally supersymmetric SYM theories in lower dimensions can be derived from the $d=10$ theory by dimensional reduction, and (\ref{SYMdsLC}) straightforwardly gives the form of $\da_{\ep\Qbar}\phi_a$ in all those cases. In particular, in the $d=4$ $N=4$ case, $\da_{\ep\Qbar}\phi_a$ is given by (\ref{d4SYMdsLC}), which reproduces the result of \cite{ABKR} where it was found through the analysis of constraints imposed by the supergroup $PSU(2,2|4)$.~\footnote{
We have also verified that the Hamiltonian, $H$, of the $d=10$ SYM is given by the quadratic form of $\da_{\ep\Qbar}\phi_a$ in (\ref{SYMdsLC}). This extends the validity of this property, discovered in \cite{ABKR}, to maximally supersymmetric theories in dimensions higher than four. Further details will be given elsewhere.
}

In the case of $d=3$ $N=8$ Bagger-Lambert-Gustavsson (BLG) theory, our result for $\da_{\epbar Q}\phi_a$ is given in (\ref{dsBLG}) together with (\ref{dsBLGfree}). As discussed further in \cite{BBKR}, the Hamiltonian of the BLG theory is also given by the quadratic form of $\da_{\epbar Q}\phi_a$. In \cite{BBKR}, we analyzed implications of the supergroup $OSp(2,2|8)$ on the structure of $\da_{\epbar Q}\phi_a$, and found one solution to a subset of constraints imposed by the supergroup. The fact that this solution matched the one derived in this paper directly from the BLG theory was then used to claim that the remaining constraints are also satisfied.

Our results complete those in \cite{BLN,BN} and establish the bridge between the covariant formulations of the SYM and BLG theories, given in \cite{BSS,GSO} and \cite{BL1,Gus1,BL2}, and the ``bottom-up'' constructions advocated in \cite{ABR1,ABKR,BBKR}. By extending the algorithm of Section \ref{sec-alg} to make it applicable to (maximally supersymmetric) supergravity theories as well, we intend to establish a similar bridge between \cite{CJS,CJ} and \cite{ABR2,BKR}, and to extend the results of \cite{ABR2,BKR} to all orders in the gravitational coupling constant $\kappa$.

\vspace{40pt}
\noindent{\bf\large Acknowledgements}
\vspace{10pt}

\noindent
I thank Jon Bagger and my collaborators on the closely related project \cite{BBKR}, Lars Brink, Sung-Soo Kim and Pierre Ramond, for helpful discussions and comments on the manuscript. I especially thank Sung-Soo Kim for helping to correct an earlier version of equation (\ref{dsAblg1}). I also thank Warren Siegel for a discussion of residual gauge invariance in the light-cone gauge. This research was supported by the Department of Energy Grant No. DE-FG02-97ER41029.

\appendix

\section{Complex conjugation}
\label{sec-cc}

In this paper, we use complex conjugation which interchanges the order of operands~\footnote{
Hermitian conjugation is defined with respect to a scalar product such as $(\Phi_1,\Phi_2)=\int\Phi_1^\ast\mc{K}\Phi_2$ where $\mc{K}$ is the integration kernel. Given an operator $\mc{A}$, its Hermitian conjugate $\mc{A}^\dagger$ is defined by $(\Phi_1,\mc{A}\Phi_2)=(\mc{A}^\dagger\Phi_1,\Phi_2)$. Our complex conjugation corresponds to Hermitian conjugation with a unit kernel, $\mc{K}=1$. However, one finds that another, $\p^{+}$-dependent kernel should instead be chosen in the LC superspace \cite{ABKR,BBKR} and for this reason we refer to conjugation used in this paper as ``complex conjugation.''
}
\bea
(\mc{O}_1\mc{O}_2\dots\mc{O}_n)^\ast=(\mc{O}_n)^\ast\dots(\mc{O}_2)^\ast(\mc{O}_1)^\ast \ .
\eea
This rule applies irrespective of whether $\mc{O}$'s are bosonic or fermionic objects. When $\mc{O}$'s are fields $\Phi$'s, out of which $k$ are fermionic and $(n-k)$ are bosonic, simple reordering gives
\bea
\label{conjF}
(\Phi_1\Phi_2\dots\Phi_n)^\ast=(-)^{k(k-1)/2}\Phi_1^\ast\Phi_2^\ast\dots\Phi_n^\ast \ .
\eea
When some of $\mc{O}$'s are operators, the rule is somewhat different.
Let $\mc{B}$ ($\mc{F}$) be a bosonic (fermionic) operator and $\phi$ ($\psi$) a bosonic (fermionic) field. The action of $\mc{B}$ ($\mc{F}$) on $\phi$ ($\psi$) is given by an (anti)commutator, and we find that
\bea
(\mc{B}\phi)^\ast &=& [\mc{B},\phi]^\ast=[\phi^\ast,\mc{B}^\ast]=-[\mc{B}^\ast,\phi^\ast]=-\mc{B}^\ast\phi^\ast \nn\\
(\mc{F}\psi)^\ast &=& \{\mc{F},\psi\}^\ast=\{\psi^\ast,\mc{F}^\ast\}=+\{\mc{F}^\ast,\psi^\ast\}=+\mc{F}^\ast\psi^\ast \ ,
\eea 
and similarly $(\mc{B}\psi)^\ast=-\mc{B}^\ast\psi^\ast$ and $(\mc{F}\phi)^\ast=-\mc{F}^\ast\phi^\ast$. For $k$ fermionic operators acting on a bosonic field, we find
\bea
(\mc{F}_1\dots\mc{F}_k\phi)^\ast &=& -(-)^{k-1}\mc{F}_1^\ast(\mc{F}_2\dots\mc{F}_k\phi)^\ast \nn\\
&=& (-)^2(-)^{k-1}(-)^{k-2}\mc{F}_1^\ast\mc{F}_2^\ast(\mc{F}_3\dots\mc{F}_k\phi)^\ast \nn\\
&=& (-)^k(-)^{k(k-1)/2}\mc{F}_1^\ast\dots\mc{F}_k^\ast\phi^\ast \ .
\eea
For $n$ operators acting on a bosonic field we then have
\bea
\label{conjOp}
(\mc{O}_1\mc{O}_2\dots\mc{O}_n\phi)^\ast=(-)^n(-)^{k(k-1)/2}(\mc{O}_1)^\ast(\mc{O}_2)^\ast\dots(\mc{O}_n)^\ast\phi^\ast \ ,
\eea
where $k\leq n$ is the number of fermionic operators. In this paper, we use the uniform convention that the result of complex conjugation of a \emph{field} with upper (lower) $SU(4)$ indices is given by the same field with lower (upper) $SU(4)$ indices. For example,
\bea
(\ta^m)^\ast=\ta_m, \quad (\chi^m)^\ast=\chi_m, \quad (C^{m n})^\ast=C_{m n}, \quad (\zeta^m)^\ast=\zeta_m \ .
\eea
(If the field carries no $SU(4)$ indices, the conjugation adds (removes) the bar, e.g. $\phi^\ast=\phibar$.) Using $(x^\mu)^\ast=x^\mu$ and $(\ta^m)^\ast=\ta_m$, as well as the basic commutation relations
\bea
[\p_\mu,x^\nu]=\da_\mu^\nu, \quad \{\p_m,\ta^n\}=\da_m^n, \quad \{\p^m,\ta_n\}=\da^m_n \ ,
\eea
we find that complex conjugation of the bosonic and fermionic derivatives gives
\bea
(\p_\mu)^\ast=-\p_\mu, \quad (\p_m)^\ast=+\p^m, \quad (\p^m)^\ast=+\p_m \ .
\eea
For the transverse bosonic derivatives defined in (\ref{trder1}) and (\ref{trder2}), we then find
\bea
(\p)^\ast=-\pbar, \quad (\pbar)^\ast=-\p; \quad
(\p^{m n})^\ast=-\p_{m n}, \quad (\p_{m n})^\ast=-\p^{m n} \ ,
\eea
whereas the definitions of $q$'s and $d$'s in (\ref{defqm}) and (\ref{defdm}) imply that
\bea
(q^m)^\ast &=& (-\p^m+\frac{i}{\sqrt2}\ta^m\p^{+})^\ast=-\p_m+\frac{i}{\sqrt2}\ta_m\p^{+}=-q_m \nn\\
(d^m)^\ast &=& (-\p^m-\frac{i}{\sqrt2}\ta^m\p^{+})^\ast=-\p_m-\frac{i}{\sqrt2}\ta_m\p^{+}=-d_m \ .
\eea
It then follows that complex conjugation of all the derivative operators used in this paper produces an extra minus sign. This effectively cancels the $(-)^n$ in (\ref{conjOp}). Combining (\ref{conjF}) with (\ref{conjOp}), we then have a mnemonic rule for complex conjugation: \emph{raise (lower) $SU(4)$ indices, add (remove) bars, add an overall minus sign if the number of fermionic objects is} $k=2,3$ mod 4. 
As an example, we have
\bea
(\zeta^m q_m\phi)^\ast=-\zeta_m q^m\phibar \ ,
\eea
which explains the minus sign in (\ref{SFkinsusy}).

\section{Fierz and other identities}
\label{sec-Fierz}

To derive the covariant formulation of the $d=10$ SYM in Section \ref{sec-SYM} and of the $d=3$ $N=8$ BLG theory in Section \ref{sec-BLG}, we needed various identities involving $d=10$ and $d=11$ gamma matrices. These are $32\times32$ matrices satisfying
\bea
\{\Ga^{M^\rr},\Ga^{N^\rr}\}=2\eta^{M^\rr N^\rr}I_{32}, \quad \eta^{M^\rr N^\rr}=(-+\dots+) \ .
\eea
Here we denoted the $d=11$ vector index by $M^\rr$. The transition from $d=11$ to $d=10$ is done by splitting $M^\rr=(M,\ast)$, which distinguishes the matrix $\Ga_\ast$ from the rest. Gamma matrices act on 32-component spinors $\ep$, $\la_a$ and $\Psi_a$, whose conjugates are defined using the charge conjugation matrix $C$,
\bea
\label{Cmat}
\epbar\equiv\ep^T C, \quad C^T=-C, \quad (\Ga^{M^\rr})^T=-C\Ga^{M^\rr} C^{-1} \ .
\eea
This implies the following flipping property for fermionic bilinears
\bea
\epbar_2\Ga_{M_1^\rr\dots M_n^\rr}\ep_1=(-)^n\epbar_1\Ga_{M_n^\rr\dots M_1^\rr}\ep_2
=(-)^{n(n+1)/2}\epbar_1\Ga_{M_1^\rr\dots M_n^\rr}\ep_2 \ ,
\eea
where 
\bea
\Ga_{(n)}\equiv
\Ga_{M_1^\rr\dots M_n^\rr}
\equiv\frac{1}{n!}\Big(\Ga_{M_1^\rr}\Ga_{M_2^\rr}\dots\Ga_{M_n^\rr}\pm(n!-1)\text{terms}\Big)
=\Ga_{[M_1^\rr\dots M_n^\rr]} \ .
\eea
Therefore,
\bea
\label{ep2ep1sym}
\epbar_2\Ga_{(n)}\ep_1 &=& +\epbar_1\Ga_{(n)}\ep_2 \quad \text{for $n=0,3$ mod 4} \nn\\
\epbar_2\Ga_{(n)}\ep_1 &=& -\epbar_1\Ga_{(n)}\ep_2 \quad \text{for $n=1,2$ mod 4} \ .
\eea
Given a complete set of $32\times32$ matrices $\mc{O}^\mc{I}$, we have the following Fierz identity
\bea
\label{F1}
\ep_2(\epbar_1\psi)=-\frac{1}{32}\sum_\mc{J}\mc{O}_\mc{J}\psi(\epbar_1\mc{O}^\mc{J}\ep_2) \quad\text{if}\quad
{\rm Tr}(\mc{O}_\mc{I}\mc{O}^\mc{J})=32\da_\mc{I}^\mc{J} \ .
\eea
Such a complete set is given by
\bea
\mc{O}_\mc{I} &=& \Big\{I_{32},\Ga_{M^\rr},i\Ga_{M^\rr N^\rr},i\Ga_{M^\rr N^\rr K^\rr},
\Ga_{M^\rr N^\rr K^\rr L^\rr},\Ga_{M^\rr N^\rr K^\rr L^\rr P^\rr}\Big\} \nn\\
\mc{O}^\mc{I} &=& \Big\{I_{32},\Ga^{M^\rr},i\Ga^{M^\rr N^\rr},i\Ga^{M^\rr N^\rr K^\rr},
\Ga^{M^\rr N^\rr K^\rr L^\rr},\Ga^{M^\rr N^\rr K^\rr L^\rr P^\rr}\Big\} \ ,
\eea
where $M^\rr<N^\rr<K^\rr<L^\rr<P^\rr$ has to be imposed to avoid overcounting. In Sections \ref{sec-SYM} and \ref{sec-BLG}, we only need the Fierz identity with $\ep_1$ and $\ep_2$ appearing antisymmetrically. Using (\ref{ep2ep1sym}) then kills $\epbar_1\Ga_{(n)}\ep_2$ terms with $n=0,3,4$, and we find
\bea
\label{F2}
\ep_2(\epbar_1\psi)-(1\lra2)
&=& -\frac{1}{16}\Big\{
\Ga_{M^\rr}\psi(\epbar_1\Ga^{M^\rr}\ep_2)
-\frac{1}{2!}\Ga_{M^\rr N^\rr}\psi(\epbar_1\Ga^{M^\rr N^\rr}\ep_2) \nn\\
&&\hspace{40pt}
+\frac{1}{5!}\Ga_{M^\rr N^\rr K^\rr L^\rr P^\rr}\psi(\epbar_1\Ga^{M^\rr N^\rr K^\rr L^\rr P^\rr}\ep_2) \Big\} \ , \qquad
\eea
\emph{without} the condition that $M^\rr<\dots<P^\rr$. 

In the SYM case, we have $M^\rr=(M,\ast)$ and require 
\bea
\Ga_\ast\ep=+\ep, \quad \Ga_\ast\psi=-\psi \ .
\eea
(Because we have there $\psi=\Ga_N D_M\la_a$ with $\Ga_\ast\la_a=+\la_a$). Then $\epbar_2\Ga_{(n)}\ep_1=0$ if $n$ is even, and we find that (\ref{F2}) reduces to
\bea
\ep_2(\epbar_1\psi)-(1\lra2) &=& -\frac{1}{16}\Big\{
2\Ga_M\psi(\epbar_1\Ga^M\ep_2) \nn\\
&&\hspace{30pt}
+\frac{1}{5!}\Ga_{M N K L P}\psi(\epbar_1\Ga^{M N K L P}\ep_2) \Big\} \ , 
\eea
which is a key identity for Section \ref{sec-SYM}.
We also found the following identities useful
\bea
&& \Ga^M\Ga^N=\Ga^{M N}+\eta^{M N}I_{32}, \quad
\Ga^M\Ga^{N K}=\Ga^{M N K}+\eta^{M N}\Ga^K-\eta^{M K}\Ga^N \nn\\[5pt]
&&\Ga^M\Ga_{(n)}\Ga_M=(-)^n(10-2n)\Ga_{(n)}, \quad 
\Ga^M\Ga_K\Ga_M=-8\Ga_K, \quad
\Ga^M\Ga_{(5)}\Ga_M=0 \nn\\[5pt]
&&\Ga^{M N}\Ga_K\Ga_N=-16\da_K^M+7\Ga_K\Ga^M, \quad
\Ga^{M N}\Ga_{(5)}\Ga_N=-\Ga_{(5)}\Ga^M \ .
\eea

In the BLG case, we have $M^\rr=(\mu,I)$, $\Ga^{\mu\nu\la}=\eps^{\mu\nu\la}\Gat$ and require 
\bea
\Gat\ep=+\ep, \quad \Gat\psi=+\psi \ .
\eea
(Because we have there $\psi=\Ga^I\Psi_a$ or $\psi=\Ga^\mu\Ga^I\Psi_a$, and $\Gat\Psi_a=-\Psi_a$). Then $\epbar_{(2)}\Ga_{(n)}\ep_1=0$ if $\Ga_{(n)}$ contains an odd number of $\Ga^I$, and after a little algebra we find that (\ref{F2}) reduces to (cf. equation (55) in \cite{BL2})
\bea
\ep_2(\epbar_1\psi)-(1\lra2) &=& -\frac{1}{16}\Big\{
2\Ga_\mu\psi(\epbar_1\Ga^\mu\ep_2)
-\Ga_{I J}\psi(\epbar_1\Ga^{I J}\ep_2) \nn\\
&&\hspace{30pt}
+\frac{1}{4!}\Ga_{I J K L}\Ga_\mu\psi(\epbar_1\Ga^{I J K L}\Ga^\mu\ep_2) \Big\} \ ,
\eea
which is a key identity for Section \ref{sec-BLG}. (In deriving (\ref{varEpsi}) we needed a version of this identity to work out $\Psi_b(\Psibar_d\Ga^I\ep)f^{b c d}{}_a$.)
We also found the following identities useful
\bea
&&\Ga^\mu\Ga^\nu=\Ga^{\mu\nu}+\eta^{\mu\nu}I_{32}, \quad
\Ga^I\Ga^J=\Ga^{I J}+\da^{I J}I_{32} \nn\\
&&
\Ga^I\Ga^{J K L}=\Ga^{I J K L}+3\da^{I [J}\Ga^{K L]}, \quad
\Ga^{I J}\Ga_J=7\Ga^I \nn\\
&&\Ga^I\Ga_{J_1\dots J_n}\Ga_I=(-)^n(8-2n)\Ga_{J_1\dots J_n}, \quad
\Ga^I\Ga_{J K L P}\Ga_I=0, \quad
\Ga^{I J K}\Ga_I=6\Ga^{J K} \nn\\
&&\Ga^\mu\Ga_{\nu_1\dots\nu_n}\Ga_\mu=(-)^n(3-2n)\Ga_{\nu_1\dots\nu_n}, \quad
\Ga^\mu\Ga_\nu\Ga_\mu=-\Ga_\nu, \quad
\Ga^{\mu\nu\rho}\Ga_\rho=\Ga^{\mu\nu} \nn\\
&& \Ga^{I J K}\Ga_{P Q R S}\Ga_I=-\Ga^J\Ga_{P Q R S}\Ga^K-(J\lra K) \nn\\
&&\Ga^{I J K}\Ga_{P Q}\Ga_I=-4\Ga_{P Q}\Ga^{J K}
+(3\Ga^J\Ga_{P Q}\Ga^K-16\da^J_P\da^K_Q-(J\lra K)) \nn\\
&& 
\Ga^{I J}\Ga_{P Q}\Ga_J=4\Ga^I\Ga_{P Q}-\Ga_{P Q}\Ga^I, \quad
\Ga^{I J}\Ga_{P Q R S}\Ga_J=-\Ga_{P Q R S}\Ga^I \ .
\eea

\section{'t Hooft symbols and $d=6$ gamma matrices}
\label{sec-gm6}

A convenient representation for $d=6$ gamma matrices can be built starting with 't Hooft symbols (see appendices in \cite{tH,BVvN})
\bea
\eta_{a m n} &=& \eps_{a m n 4}+\da_{a m}\da_{n 4}-\da_{a n}\da_{m 4} \nn\\
\wt\eta_{a m n} &=& \eps_{a m n 4}-\da_{a m}\da_{n 4}+\da_{a n}\da_{m 4} \ ,
\eea
where $a=1,2,3$ (only in this and the next appendix) and $m=1,2,3,4$. For each $a$, they are (real) $4\times4$ matrices. Explicitly, (cf. appendices in \cite{DHEG,Westra})
\bea
&& \eta_1=+\si_1\otimes i\si_2, \quad
\eta_2=-\si_3\otimes i\si_2, \quad
\eta_3=i\si_2\otimes I_2 \nn\\
&& \wt\eta_1=-i\si_2\otimes\si_1, \quad
\wt\eta_2=-I_2\otimes i\si_2, \quad
\wt\eta_3=i\si_2\otimes\si_3 \ ,
\eea
where the matrix on the left of ``$\otimes$'' multiplies each element of the matrix on the right of it, and $\si_a$ are standard Pauli matrices so that
\bea
I_2=\bpm 1 & 0 \\ 0 & 1 \epm, \quad
\si_1=\bpm 0 & 1 \\ 1 & 0 \epm, \quad
i\si_2=\bpm 0 & 1 \\ -1 & 0 \epm, \quad
\si_3=\bpm 1 & 0 \\ 0 & -1 \epm .
\eea
Using $\si_a\si_b=\da_{a b}+i\eps_{a b c}\si_c$, the following properties of 't Hooft symbols can be established
\bea
&&\eta_{a m k}\eta_{b n k}=\da_{a b}\da_{m n}+\eps_{a b c}\eta_{c m n}, \quad
\eta_{a m k}\wt\eta_{b n k}=\eta_{a n k}\wt\eta_{b m k}
\nn\\
&& \wt\eta_{a m k}\wt\eta_{b n k}=\da_{a b}\da_{m n}+\eps_{a b c}\wt\eta_{c m n}, \quad
\eta_1\eta_2\eta_3=\wt\eta_1\wt\eta_2\wt\eta_3=I_4
\nn\\
&&\eta_{a m n}\eta_{b m n}=\wt\eta_{a m n}\wt\eta_{b m n}=4\da_{a b}, \quad
\eta_{a m n}\wt\eta_{b m n}=0
\nn\\
&&\eta_{a m n}=+\half\eps_{m n k l}\eta_{a k l}, \quad
\eta_{a m n}\eta_{a k l}=\da_{m k}\da_{n l}-\da_{m l}\da_{n k}+\eps_{m n k l}
\nn\\
&&\wt\eta_{a m n}=-\half\eps_{m n k l}\wt\eta_{a k l}, \quad
\wt\eta_{a m n}\wt\eta_{a k l}=\da_{m k}\da_{n l}-\da_{m l}\da_{n k}-\eps_{m n k l} \ .
\eea
Defining now the following complexified objects
\bea
\Si_{\wh I m n}=\eta_{a m n}\da^{\wh I}_a+i\wt\eta_{a m n}\da^{\wh I}_{a+3}, \quad
\Si^{\wh I m n}=\eta_{a m n}\da^{\wh I}_a-i\wt\eta_{a m n}\da^{\wh I}_{a+3} \ ,
\eea
where $\wh I=1,2,3,4,5,6$, one can easily prove that
\bea
&&\Si^{\wh I m n}=(\Si_{\wh I m n})^\ast=\half\eps^{m n k l}\Si_{\wh I k l}, \quad
\Si_{\wh I m n}=-\Si_{\wh I n m} \nn\\
&&\Si_{\wh I m n}\Si^{\wh I k l}=2(\da_m^k\da_n^l-\da_m^l\da_n^k), \quad
\Si_{\wh I m n}\Si_{\wh I k l}=2\eps_{m n k l}, \quad
\Si_{\wh I m n}\Si^{\wh J m n}=4\da_{\wh I \wh J} \nn\\
&&\Si_{\wh I m k}\Si^{\wh J k n}+\Si_{\wh J m k}\Si^{\wh I k n}=-2\da_{\wh I\wh J}\da_m^n 
\qrq \Si_{\wh I}\Si^{\wh J}+\Si_{\wh J}\Si^{\wh I}=-2\da_{\wh I\wh J}I_4
\nn\\
&&\Si_{1 m k}\Si^{2 k l}\Si_{3 l p}\Si^{4 p q}\Si_{5 q r}\Si^{6 r n}=-i\da_m^n
\qrq \Si_1\Si^2\Si_3\Si^4\Si_5\Si^6=-i I_4 \ . \nn\\
\eea
The last two properties guarantee that defining
\bea
\wh\Ga^{\wh I} = \bpm 0 & \Si^{\wh I m n} \\ \Si_{\wh I m n} & 0 \epm \ ,
\eea
we find that these $8\times8$ matrices satisfy
\bea
\{\wh\Ga^{\wh I},\wh\Ga^{\wh J}\}=-2\da_{\wh I\wh J}I_8, \quad
\wh\Ga^1\wh\Ga^2\wh\Ga^3\wh\Ga^4\wh\Ga^5\wh\Ga^6=i\bpm I_4 & 0 \\ 0 & -I_4 \epm \ ,
\eea
so that $\wh\Ga^{\wh I}$ form a representation of the $d=6$ Clifford algebra.

\section{A representation for $d=11$ gamma matrices}
\label{sec-gm11}

Given the above representation for the $d=6$ ($8\times8$) gamma matrices $\wh\Ga^{\wh I}$, $\wh I=1,2,3,4,5,6$, and the following representation for the $d=4$ ($4\times4$) gamma matrices $\ga^{\mu^\rr}$, $\mu^\rr=(0,a)$, $a=1,2,3$,
\bea
\ga^0=\bpm 0 & I_2 \\ I_2 & 0 \epm, \quad
\ga^a=\bpm 0 & \si^a \\ -\si^a & 0 \epm; \quad
\ga_5=i\ga^0\ga^1\ga^2\ga^3=\bpm -I_2 & 0 \\ 0 & I_2 \epm ,\;
\eea
we choose $d=11$ ($32\times32$) gamma matrices $\Ga^{M^\rr}$, $M^\rr=0,\dots,10$, as follows
\bea
\Ga^{\mu^\rr}=i\ga^{\mu^\rr}\otimes I_8, \quad
\Ga^I=i\ga_5\otimes\wh\Ga^{I-3}, \quad
\Ga^{10}=\Ga_\ast\equiv\Ga_0\Ga_1\dots\Ga_9 \ ,
\eea
where $I=4,\dots,9$. We have
\bea
\{\ga^{\mu^\rr},\ga^{\nu^\rr}\}=-2\eta^{\mu^\rr\nu^\rr}I_4, \quad
\{\Ga^{M^\rr},\Ga^{N^\rr}\}=+2\eta^{M^\rr N^\rr}I_{32} \ ,
\eea
where $\eta$'s have signature $(-+\dots+)$. We also find that
\bea
\Ga_\ast=-\Ga^0\dots\Ga^9=\ga^0\ga^1\ga^2\ga^3\otimes\wh\Ga^1\wh\Ga^2\wh\Ga^3\wh\Ga^4\wh\Ga^5\wh\Ga^6
=\ga_5\otimes\bpm I_4 & 0 \\ 0 & -I_4 \epm \ .
\eea
As the charge conjugation matrix $C$ satisfying (\ref{Cmat}), we take
\bea
C=i C_4\otimes\bpm 0 & I_4 \\ I_4 & 0 \epm, \quad
C_4=\bpm i\si_2 & 0 \\ 0 & -i\si_2 \epm \ ,
\eea
where $C_4$ satisfies $C_4^T=-C_4$ and $C_4\ga^{\mu^\rr}C_4^{-1}=-(\ga^{\mu^\rr})^T$.

\section{Various forms of the Fundamental Identity}
\label{sec-FI}

In Section \ref{sec-BLG}, we found that closure of the supersymmetry algebra in the BLG theory requires a Jacobi-like identity (\ref{FI}) for the structure constants. Using that
\bea
4f^{[a b c}{}_g f^{e]f g}{}_d=f^{a b c}{}_g f^{e f g}{}_d-3f^{e[a b}{}_g f^{c]f g}{}_d \ ,
\eea
we see that (\ref{FI}) is equivalent to
\bea
\label{FI2}
f^{a b c}{}_g f^{e f g}{}_d=3f^{e[a b}{}_g f^{c]f g}{}_d \ .
\eea
Using the (totally antisymmetric) triple product in (\ref{triple}), this identity can be stated as
\bea
\label{FI2a}
[\al,\beta,[X,Y,Z]]=[X,\beta,[\al,Y,Z]]+[Y,\beta,[\al,Z,X]]+[Z,\beta,[\al,X,Y]] \ .
\eea
Applying this identity to the terms on the right hand side in the following fashion
\bea
[\beta,X,[\al,Y,Z]] &=& [\al,X,[\beta,Y,Z]]+[Y,X,[\beta,Z,\al]]+[Z,X,[\beta,\al,Y]] \nn\\{}
[\beta,Y,[\al,Z,X]] &=& [\al,Y,[\beta,Z,X]]+[Z,Y,[\beta,X,\al]]+[X,Y,[\beta,\al,Z]] \nn\\{}
[\beta,Z,[\al,X,Y]] &=& [\al,Z,[\beta,X,Y]]+[X,Z,[\beta,Y,\al]]+[Y,Z,[\beta,\al,X]] \ , \nn\\ 
\eea
and summing this column by column (the sum of the first column is minus the sum of the second one, while the sum of the third column equals the sum of the fourth one), we find that
\bea
\label{FI3a}
[\al,\beta,[X,Y,Z]]=[X,Y,[\al,\beta,Z]]+[Y,Z,[\al,\beta,X]]+[Z,X,[\al,\beta,Y]] \ ,
\eea
which is the Fundamental Identity as given in equation (9) of \cite{BL2}, and which can be equivalently stated as
\bea
\label{FI3}
f^{a b c}{}_g f^{e f g}{}_d=3f^{e f[a}{}_g f^{b c]g}{}_d \ .
\eea
Applying (\ref{FI3a}) to the terms on its right hand side in the following fashion
\bea
[X,Y,[\al,\beta,Z]] &=& [\al,\beta,[X,Y,Z]]+[\beta,Z,[X,Y,\al]]+[Z,\al,[X,Y,\beta]] \nn\\{}
[Y,Z,[\al,\beta,X]] &=& [\al,\beta,[Y,Z,X]]+[\beta,X,[Y,Z,\al]]+[X,\al,[Y,Z,\beta]] \nn\\{}
[Z,X,[\al,\beta,Y]] &=& [\al,\beta,[Z,X,Y]]+[\beta,Y,[Z,X,\al]]+[Y,\al,[Z,X,\beta]] \ , \nn\\
\eea
and summing this column by column, we find that
\bea
[\al,\beta,[X,Y,Z]]=\half\Big([X,\beta,[\al,Y,Z]]+\text{cycle}(X,Y,Z)-(\al\lra\beta)\Big) \ .
\eea
This agrees with (\ref{FI2a}), but is not equivalent to it. Namely, we recover only the part of (\ref{FI2a}) which is antisymmetric in $\al$ and $\beta$, but not the symmetric part:
\bea
0=\half\Big([X,\beta,[\al,Y,Z]]+\text{cycle}(X,Y,Z)+(\al\lra\beta)\Big) \ .
\eea
Therefore, we conclude that (\ref{FI2a}) is \emph{stronger} than (\ref{FI3a}). Equivalently, (\ref{FI2}) is stronger than (\ref{FI3}). (We differ on this point with \cite{GNP,Gus2,BL4}.)~\footnote{
When the (Killing) metric for gauge indices is introduced, and $f^{a b c d}$ is totally antisymmetric in all four indices, the two forms of the Fundamental Identity, equations (\ref{FI2}) and (\ref{FI3}), become equivalent. Indeed, (\ref{FI3}) then becomes $f^{g a b c}f^{e f d g}=3f^{g e f[a}f^{b c]d g}$ and this implies (\ref{FI2}) in the form $f^{g a b c}f^{e f d g}=3f^{g e[a b}f^{c]f d g}$ after simply switching the two $f$'s and relabeling the indices.
}

Finally, we note that (\ref{FI3}) can also be stated as follows
\bea
\label{FI4}
-f^{c d b}{}_g f^{e f g}{}_a+f^{e f b}{}_g f^{c d g}{}_a=
 f^{e f c}{}_g f^{d g b}{}_a-f^{e f d}{}_g f^{c g b}{}_a \ .
\eea
This form is used in the next appendix.

\section{Adding and removing the tilde}
\label{sec-tilde}

In Section \ref{sec-BLG}, we introduced $\wt\La^b{}_a$ and $\wt A_\mu{}^b{}_a$ but did not give any special meaning to the tilde. Let us now follow \cite{BL2} and write
\bea
\boxed{
\wt\La^b{}_a=\La_{c d}f^{c d b}{}_a \ , \quad
\wt A_\mu{}^b{}_a=A_{\mu c d}f^{c d b}{}_a \ ,
}
\eea
which \emph{defines} $\La_{c d}$ and $A_{\mu c d}$. We take $\La_{c d}$ and $A_{\mu c d}$ to be antisymmetric in the gauge indices (as $f^{c d b}{}_a$ kills the symmetric part anyway). Writing now (\ref{AgtrBLG}) as follows
\bea
f^{c d b}{}_a\da_\La A_{\mu c d} \ =\ f^{c d b}{}_a\p_\mu\La_{c d}
&-& (f^{c d b}{}_g\La_{c d})(f^{e f g}{}_a A_{\mu e f}) \nn\\
&+& (f^{e f b}{}_g A_{\mu e f})(f^{c d g}{}_a\La_{c d}) \ ,
\eea
and using the Fundamental Identity in the form (\ref{FI4}), we find
\bea
f^{c d b}{}_a(\da_\La A_{\mu c d}-\p_\mu\La_{c d})
&=& 2f^{e f c}{}_g f^{d g b}{}_a A_{\mu e f}\La_{c d} \nn\\
&=& 2f^{e f g}{}_d f^{c d b}{}_a A_{\mu e f}\La_{g c} \ .
\eea
Removing $f^{c d b}{}_a$ while maintaining the $[c d]$ symmetry then yields
\bea
\boxed{
\da_\La A_{\mu c d}=\p_\mu\La_{c d}+f^{e f g}{}_d A_{\mu e f}\La_{g c}-f^{e f g}{}_c A_{\mu e f}\La_{g d} \ ,
}
\eea
which closely resembles (\ref{AgtrSYM}). From (\ref{BLGsusy}) it also trivially follows that
\bea
\da_\ep A_{\mu c d}=\frac{i}{2}\epbar\Ga_\mu\Ga^I X^I_c\Psi_d-(c\lra d) \ ,
\eea
and therefore one could take $A_{\mu c d}$ to be \emph{the} gauge field in the BLG theory.

Conversely, one could introduce the tilde for the Yang-Mills gauge field $A_{M a}$. Multiplying the gauge transformation (\ref{AgtrSYM})
\bea
\da_\La A_{M c}=\p_M\La_c+f^{e f}{}_c A_{M e}\La_f 
\eea
with $f^{b c}{}_a$ and using the Jacobi identity (\ref{Jacobi}) in the form
\bea
f^{e f}{}_c f^{b c}{}_a=-f^{f b}{}_c f^{e c}{}_a-f^{b e}{}_c f^{f c}{}_a \ ,
\eea
we find
\bea
f^{b c}{}_a\da_\La A_{M c}=f^{b c}{}_a\p_\mu\La_c
-f^{b f}{}_c f^{c e}{}_a A_{M e}\La_f
+f^{b e}{}_c f^{c f}{}_a A_{M e}\La_f \ .
\eea
Defining now
\bea
\boxed{
\wt\La^b{}_a=\La_c f^{b c}{}_a, \quad
\wt A_M{}^b{}_a=A_{M c}f^{b c}{}_a
}
\eea
yields
\bea
\boxed{
\da_\La \wt A_M{}^b{}_a=\p_M\wt\La^b{}_a-\wt\La^b{}_c\wt A_M{}^c{}_a+\wt A_M{}^b{}_c\wt\La^c{}_a 
}
\eea
which \emph{coincides} with (\ref{AgtrBLG}). The fact that one needs to use the Jacobi identity (or the Fundamental Identity) to go between the two (with and without tilde) formulations, explains why in the SYM case (Section \ref{sec-SYM}) we needed the Jacobi identity to prove that the covariant derivative transforms covariantly, while this was automatic in the BLG case (Section \ref{sec-BLG}).



\end{document}